\documentclass[preprint]{aastex}
\usepackage{emulateapj5}
\usepackage{lambdabar}
\usepackage{amsmath}
\usepackage{amssymb}

\newcommand{\lambdaC}{\lambdabar_C}
\newcommand{\lnlambda}{\ln \Lambda}
\newcommand{\psiinf}{\psi_\infty}
\newcommand{\Zinf}{Z_\infty}
\newcommand{\alphaF}{\alpha_F}
\newcommand{\eb}{\epsilon_B}

\newcommand{\sci}[2]{\mbox{$#1 \times 10^{#2}$}}

\newcommand{\uvc}[1]{\mbox{\boldmath ${\hat{#1}}$}} 
\newcommand{\vc}[1]{\mbox{\boldmath ${\bf #1}$}}    

\newcommand{\average}[1]{\langle #1 \rangle}

\shorttitle{Pair-production multiplicities}
\shortauthors{Hibschman and Arons}

\begin{document}
\title{Pair-production multiplicities in rotation-powered pulsars}
\author{Johann A.\ Hibschman \altaffilmark{1,2}}
\and
\author{Jonathan Arons \altaffilmark{1,2,3}}
\affil{University of California, Berkeley \altaffilmark{4}}
\altaffiltext{1}{Theoretical Astrophysics Center}
\altaffiltext{2}{Physics Department}
\altaffiltext{3}{Astronomy Department}
\altaffiltext{4}{Address correspondence to J.~A. Hibschman, Astronomy 
Dept., 601 Campbell Hall, U.C., Berkeley 94720-3411; email:
{\tt jhibschman@astron.berkeley.edu}}

\begin{abstract}
  We discuss the creation of electron-positron cascades in the context
  of pulsar polar cap acceleration models and derive several useful
  analytic and semi-analytic results for the spatial extent and energy
  response of the cascade. Instead of Monte Carlo simulations, we use
  an integro-differential equation which describes the development of the
  cascade energy spectrum in one space dimension quite well, when it is
  compared to existing Monte Carlo models.  We reduce
  this full equation to a single integral equation, from which we can derive
  useful results, such as the energy loss between successive generations
  of photons and the spectral index of the response.
  
  We find that a simple analytic formula represents the pair cascade
  multiplicity quite well, provided that the magnetic field is below $10^{12}$
  Gauss, and that an only slightly more complex formula matches the
  numerically-calculated cascade at all other field strengths.
  
  Using these results, we find that cascades triggered by gamma rays
  emitted through inverse Compton scattering of thermal photons from the
  neutron star's surface, both
  resonant and non-resonant, are important for the
  dynamics of the polar cap region in many pulsars. In these objects,
  the expected multiplicity of pairs generated by a
  single input particle is lower than previously found in cascades
  initiated by curvature emission, frequently being on the order of 10
  rather than  $\sim 1000$ as usually quoted.  Such pulsars also are
  expected to be less luminous in polar-cap gamma rays than when
  curvature emission triggers the cascade, a topic which will be the
  subject of a subsequent paper.
\end{abstract}


\maketitle

\section{Introduction}

The polar-cap pair-production model of pulsar emission has remained
the chief theory of the region thought to give rise to the radiative
emission for the past 30 years.  Particles accelerate from the surface
of the neutron star, drawn by the induced fields of the rotating
magnetosphere.  As first discussed by \citet{sturrock71}, these
particles emit $\gamma$-rays which pair-produce in the high
background magnetic field.   These pairs then short out the magnetic
field \citep{ruderman75}, preventing further acceleration.

The physics of this pair production process has been partially explored,
with progress ranging from the initial generational models of
\citet{tademaru73}, the pair formation front structure calculations
of \citet{arons79,arons81}, 
the detailed Monte Carlo simulations of \citet{daugherty82}, and the
more recent ``full-cascade'' generational model of \citet{zhang00}.
Our paper describes an improved
analytical and theoretical understanding of this pair-production
mechanism, not only reproducing the final $\gamma$-ray and pair spectra
calculated by \citet{daugherty82}, but detailing the spatial extent
of the pair-production process itself, giving a more thorough
understanding of the process.

These straightforward analytic approximations to the pair production
process were the basis of the model of \cite{hibschman01},
henceforth called Paper I.  In this paper, we exhibit the full shower
theory underlying those approximations and find a variety of formulae
which represent the numerical solutions of the cascade equations, thus
allowing users of the cascade physics to easily apply the results to other
problems.  We give a full description of the pair plasma emergent from
pulsar polar caps. We reserve discussion of the emergent $\gamma$-ray
luminosities and spectra to a separate paper, as these are of direct
observational relevance.

\section{Gamma-ray opacity}

The dominant opacity for pair production over pulsar polar caps is
the $\gamma$-$B$ process, wherein a high-energy photon interacts with
the background magnetic field to form an electron-positron pair.
The competing process of $\gamma$-$\gamma$ pair production, wherein a
high-energy photon interacts with a background X-ray to form a pair,
is only competitive if the background X-ray flux is equivalent to a
black-body with temperature nearly $10^7$ K, well beyond the observed
limits on stellar temperatures.

According to \citet{erber66}, the opacity for $\gamma$-$B$
pair-production is
\begin{equation}
  \alpha_B(\epsilon, \psi) = 0.23 \frac{\alpha_F}{\lambdaC} \frac{B}{B_q}
             \sin \psi \exp \left( - \frac{8}{3 \chi} \right)
  \label{eq:gB-opacity}
\end{equation}
where $B_q$ is the critical quantum magnetic field, $B_q = e/\alpha 
\lambdaC^2 = \sci{4.41}{13}$ Gauss, $\alpha_F = e^2/\hbar c \approx 1/
137$ is the fine structure constant, $\lambdaC = \hbar/m c = 
\sci{3.86}{-11}$ cm is the reduced Compton wavelength, $B$ is the local 
magnetic field strength, $\psi$ is the pitch angle between the photon 
momentum and the local magnetic field, and
\begin{equation}
  \chi \equiv \epsilon \frac{B}{B_q} \sin \psi.
\end{equation}
where $\epsilon$ is the photon energy, in units of $mc^2$. For the 
remainder of this paper, all energies will be quoted in terms of $mc^2$, 
for convenience.

This expression is accurate provided that $\epsilon \sin \psi > 1$, so
that the created pair is in a high Landau level, and provided that $B$
is small compared to $B_q$; $B < B_q/3$ suffices.  For pair production
into low Landau levels, the full cross section must be used
\citep{daugherty83}, and for magnetic fields
equal to or higher than $B_q$, higher-order corrections become
important, as discussed in \citet{harding97}.  Most pulsars, however,
pair produce into high Landau levels for the most significant photons
and have $B < B_q/3$.  As this work intends to describe the ``typical''
pulsar, we neglect these high-energy and high-field effects.

\subsection{Optical depth}

Because the relativistic primary electron beam follows the magnetic
field, the primary photons are beamed close to parallel to the local
field.  If the beam electrons have a Lorentz factor $\gamma$, the
primary photons are beamed into a cone of opening angle $\psi \sim
1/\gamma$.  Typical beam Lorentz factors are of order $10^3$ or
higher, giving an initial pitch angle small compared to the pitch
angles required for pair-production.  To a good approximation, then,
we can treat the photons as if they are injected precisely parallel to
the field lines.

As the photons propagate through the magnetosphere, the pitch angle
between the photon momentum and the background magnetic field steadily
increases.  This pitch angle is given by $\sin \psi =
|\uvc{k}\times\uvc{B}(\vc{r})|$, where \uvc{k} is the photon momentum
direction, and \uvc{B} is the local magnetic field direction.  If we
assume $\psi$ small enough that $\sin \psi \approx \psi$, the change
in $\psi$ along the photon path is then $d\psi/ds = |\uvc{k} \times
(\uvc{k} \cdot \nabla ) \uvc{B}(r)|$.

Since $\psi$ is small, \uvc{k} remains close to \uvc{B}, so we can
substitute, yielding $d\psi/ds \approx |\uvc{B} \times (\uvc{B} \cdot
\nabla) \uvc{B}(\vc{r})| = \rho(\vc{r})^{-1}$, where $\rho(\vc{r})$ is
the local magnetic field radius of curvature.  This is accurate to
second order in $\psi$.

For a dipole field, the photon pitch angle depends on radius (through
second order in the colatitude, $\theta$) as
\begin{equation}
  \psi = \psiinf \left( 1 - \frac{r_e}{r} \right)
  \label{eq:psi_of_r}
\end{equation}
where $r_e$ is the radius of emission and $r$ is the current radius,
both measured from the center of the dipole, and $\psiinf \equiv
r_e/\rho_e$, where $\rho_e =(4/3) (R_* /\theta_*) f_\rho (r_e/R_*)^{3/2}$
is the field line radius of
the magnetic colatitude of the field line in question at the surface
of the star and $f_\rho \leq 1$ is a factor used to crudely take into
account the
possibility of non-dipolar components to the magnetic field at low
altitude.   Numerically this is $\psiinf = 0.011
(\theta_*/\theta_C) P^{-1/2} (r/R_*)^{1/2} f_\rho$, with $\theta_C$
the colatitude of the last closed field line, $\theta_C = \sqrt{R_*/R_L}$
and $P$ is the period in seconds.  If the dipole moment is displaced
close to the surface of the star, GR bending of the photon paths
would change this, but for almost all other cases GR perturbations are
negligible.

The magnetic field and radius of curvature may then be rewritten as
\begin{eqnarray}
  B & = & B_e \left( \frac{r}{r_e} \right)^{-3} = B_e \left( 1 -
     \frac{\psi}{\psiinf} \right)^3 \\
  \rho & = & \rho_e \left( \frac{r}{r_e} \right)^{1/2} = \rho_e \left( 1 -
     \frac{\psi}{\psiinf} \right)^{-1/2}
\end{eqnarray}
where $B_e$ and $\rho_e$ are respectively the magnetic field and radius of
curvature at the emission point.

Using these relations to write the optical depth using $t \equiv
\psi/\psiinf$ as the integration variable, letting $\sin \psi \approx
\psi$, yields
\begin{equation}
  \tau(\psi) =  0.23 \frac{\alpha \rho_e}{\lambdaC}
    \frac{B_e}{B_q} \psiinf^2 \int_0^{\psi/\psiinf} t(1-t)^{5/2} \exp
    \left[ - \frac{\Zinf}{t(1-t)^3} \right] dt
  \label{eq:rawtau}
\end{equation}
where
\begin{equation}
  \Zinf \equiv \frac{8}{3} \frac{B_q}{\epsilon B_e
    \psiinf}
\end{equation}
and $\epsilon$ is the photon energy in units of $mc^2$.

The exponential above has a maximum at $t=1/4$.  Physically, this is
the point where the increase in opacity due to the increasing pitch
angle balances the decrease in opacity due to the steadily declining
magnetic field.

Provided that $\Zinf$ is not too small, the integrand is a sharply
peaked function of $t$, and can therefore be integrated using the steepest
descents method, yielding
\begin{equation}
  \tau_\infty (\epsilon) = 0.23 \frac{\alpha \rho}{\lambdaC}
    \frac{B}{B_q} \psiinf^2 \frac{3^4}{2^{14}}
    \sqrt{\frac{6\pi}{\Zinf (\epsilon)}}
    \exp \left[ -\frac{256}{27} \Zinf (\epsilon) \right].
  \label{eq:tau_infty}
\end{equation}
When compared to numerical calculations, this expression is accurate
to better than 10\% in the neighborhood of $\tau=1$, for the range of
magnetic fields and radii of curvature found in pulsars.

Photons with $\tau_\infty < 1$ will escape the magnetosphere, while
those with $\tau_\infty > 1$ will be absorbed.  If $\tau_\infty
(\epsilon)$ is large, most of the absorption takes place on the
leading edge of the exponential.  Since the opacity is increasing
exponentially in that regime, we can approximate the integral by
expanding around the upper endpoint to find
\begin{equation}
   \tau(\epsilon, t) \approx \Lambda_1 t^3 e^{-\Zinf (\epsilon) f(t)}
   \label{eq:tau}
\end{equation}
where
\begin{equation}
  \Lambda_1 = 0.086 \frac{\alpha \rho_e}{\lambdaC} \left(
    \frac{B_e}{B_q} \right)^2 \epsilon \psiinf^3
\end{equation}
and where $f(t)$ has been defined to be $t^{-1} (1-t)^{-3}$ for
$t<1/4$ and 256/27 for $t>1/4$.  (This result was previously derived
in \citet{arons79}.)  Since the opacity decreases rapidly after
$t=1/4$, the optical depth saturates at that point.

\subsection{Absorption peak}
\label{sec:absorption_peak}

Given this opacity, we can find where any given photon will be
absorbed.  Due to the exponential dependence of the opacity, all
photons are effectively absorbed at the maximum of $\alpha
\exp(-\tau)$, or approximately at $\tau = 1$.  Using (\ref{eq:tau})
for $\tau$, we find this peak occurs at
\begin{equation}
  t_a (1 - t_a)^3 = \frac{8}{3} \frac{B_q}{\epsilon \psiinf B \lnlambda}
\end{equation}
where $\lnlambda \equiv \ln (\Lambda_1(\epsilon) t^3)$.  In the limit
of small $t_a$, this becomes
\begin{equation}
  \psi_a = \frac{8}{3} \frac{B_q}{\epsilon B \lnlambda}.
  \label{eq:psi_a}
\end{equation}
Since the opacity saturates past $\psi_a = \psiinf/4$, the minimum
photon energy required to pair produce is
\begin{equation}
  \epsilon_a = \frac{32}{3} \frac{B_q}{B \psiinf} \frac{1}{\lnlambda}
   \; mc^2.
\end{equation}
This energy should be thought of as a critical scaling energy, not as
the actual minimum energy which will pair produce, since the $(1-t)$
term was neglected.  The actual minimum energy which will pair produce
is
\begin{equation}
  \epsilon_{min} = \frac{64}{27} \epsilon_{a}.
\end{equation}

As long as $\epsilon \gtrsim 5 \epsilon_a$, the small-$t$ limit
(\ref{eq:psi_a}) is appropriate, and we find that a photon will be
absorbed after propagating
\begin{equation}
  \Delta r = \frac{1}{4} \frac{\epsilon_a}{\epsilon} r_e.
  \label{eq:dr_a}
\end{equation}

These expressions can be made tractable by treating $\lnlambda$ as a
constant.  Self-consistently evaluating $\lnlambda$ at $t=1/8$ then
yields \citep{arons79}.
\begin{equation}
  \lnlambda = 16.2 + \ln B_{12} - \frac{1}{2} \ln P.
\end{equation}

\section{Synchrotron emission}
\label{sec:synchrotron}

A charged particle with pitch angle $\psi$ with respect to the
magnetic field emits synchrotron radiation and rapidly spirals down to
its lowest Landau level.

The basic rate of synchrotron emission is
\begin{equation}
  \frac{\partial N_\epsilon}{\partial t} =
    \frac{\sqrt{3}}{2 \pi} \frac{\alpha c}{\lambdaC}
    \frac{\epsilon_B}{\epsilon_s} \sin \psi
    \int_{\epsilon/\epsilon_s}^\infty K_{5/3}(x)dx.
  \label{eq:Nededt}
\end{equation}
where $\epsilon_B \equiv B/B_q$, $K_{5/3}$ is the modified Bessel function
of order $5/3$, and $\epsilon_s$ is the characteristic synchrotron energy,
\begin{equation}
  \epsilon_s   =  \frac{3}{2} \epsilon_B \gamma^2 \sin \psi \; mc^2
\end{equation}
This corresponds to a total power loss of \citep{jackson75}
\begin{equation}
  P_s = \frac{2}{3} \frac{\alpha c}{\lambdaC} \epsilon_B^2 \gamma^2
    \sin^2 \psi \; mc^2 s^{-1}.
  \label{eq:synch_power}
\end{equation}

Due to synchrotron radiation, the Lorentz factor of the particle
decreases according to $\dot{\gamma} = -P_s$.  This energy loss occurs
over a distance of \sci{1.8}{-6} $\gamma$ cm, effectively
instantaneous compared to stellar scales, where we have used equation
(\ref{eq:a}) self-consistently to fix $\gamma \psi$ and let
$\sin \psi \approx \psi$.  The final
Lorentz factor may be found by noticing that the parallel component of
the particle's velocity is conserved, as can be seen by transforming
into a comoving frame.  If the initial particle has a Lorentz factor
of $\gamma_i = (1-\beta_i^2)^{-1/2}$ and is moving at an angle $\psi$
with respect to the magnetic field, then the final Lorentz factor is
\begin{equation}
  \gamma_f = \frac{1}{\sqrt{1 - \beta_i^2 \cos^2 \psi}}
    \approx \frac{\gamma_i}{\sqrt{1 + \gamma_i^2 \psi^2}}
  \label{eq:gammaf}
\end{equation}

Integrating (\ref{eq:Nededt}) over all time gives the number of
photons generated during the decay.  Using (\ref{eq:synch_power}) to
change the variable of integration from time to $\gamma$ gives

\begin{eqnarray}
  N_\epsilon(\gamma_i, \psi) & = &
    \frac{3 \sqrt{3}}{8 \pi} (\epsilon_B \sin \psi)^{-1} \times \nonumber \\
    & & \int_{\gamma_f}^{\gamma_i} \frac{d\gamma}{\gamma^2}
      \epsilon_s(\gamma)^{-1}
    \int_{\epsilon/\epsilon_s}^\infty dx \, K_{5/3}(x).
    \label{eq:Ne}
\end{eqnarray}

Changing variables to $z = \epsilon/\epsilon_s(\gamma)$ and
re-arranging gives
\begin{equation}
  N_\epsilon(\gamma_i, \psi) = \frac{3}{4 \sqrt{2} \pi} (\epsilon_B \sin
    \psi)^{-1/2} \epsilon^{-3/2} [ F(z_i) - F(z_f) ]
    \label{eq:n}
\end{equation}
where
\begin{eqnarray}
  F(t) & \equiv & \frac{3}{2} \int_{t}^\infty dz \, z^{1/2}
    \int_z^\infty dx \, K_{5/3}(x) \nonumber \\
    & = &
    \int_t^\infty dx \, K_{5/3}(x) \left( x^{3/2} -
    t^{3/2} \right)
    \label{eq:f}
\end{eqnarray}

Equation (\ref{eq:n}) gives the total number of photons of energy
$\epsilon$ emitted by a particle with initial Lorentz factor
$\gamma_i$ and initial pitch angle $\psi$.  Nearly all of these
photons are emitted in a cone with opening angle $\psi$, because the
Lorentz factor of the particle remains high, so the angle $\psi$ only
changes once almost all of the particle energy has been radiated away.

In the case of a particle created by $\gamma$-$B$ pair production from
a photon of initial energy $\epsilon_i$, $\gamma_i = \epsilon_i/2$ and
$\psi = \psi_a(\epsilon_i)$.  Defining
\begin{equation}
  a \equiv \gamma_i \psi_a = \frac{4}{3 \lnlambda}
    \frac{1}{\epsilon_B}
  \label{eq:a}
\end{equation}
we find $z_i = \lnlambda \epsilon/\epsilon_i$ and $z_f = (1+a^2)
\lnlambda \epsilon / \epsilon_i$.

With these substitutions and counting the emission from both generated
particles, the total number of synchrotron photons produced by one
incident photon pair producing at $\psi=\psi_a$ is
\begin{equation}
  N_\epsilon(\epsilon_i) = \frac{3\sqrt{3}}{8 \pi} \sqrt{\lnlambda}
    \left(\frac{\epsilon}{\epsilon_i}\right)^{-3/2} \epsilon_i^{-1}
    [F(z_i) - F(z_f)]
  \label{eq:synch_n}
\end{equation}

\begin{figure*}
  \plottwo{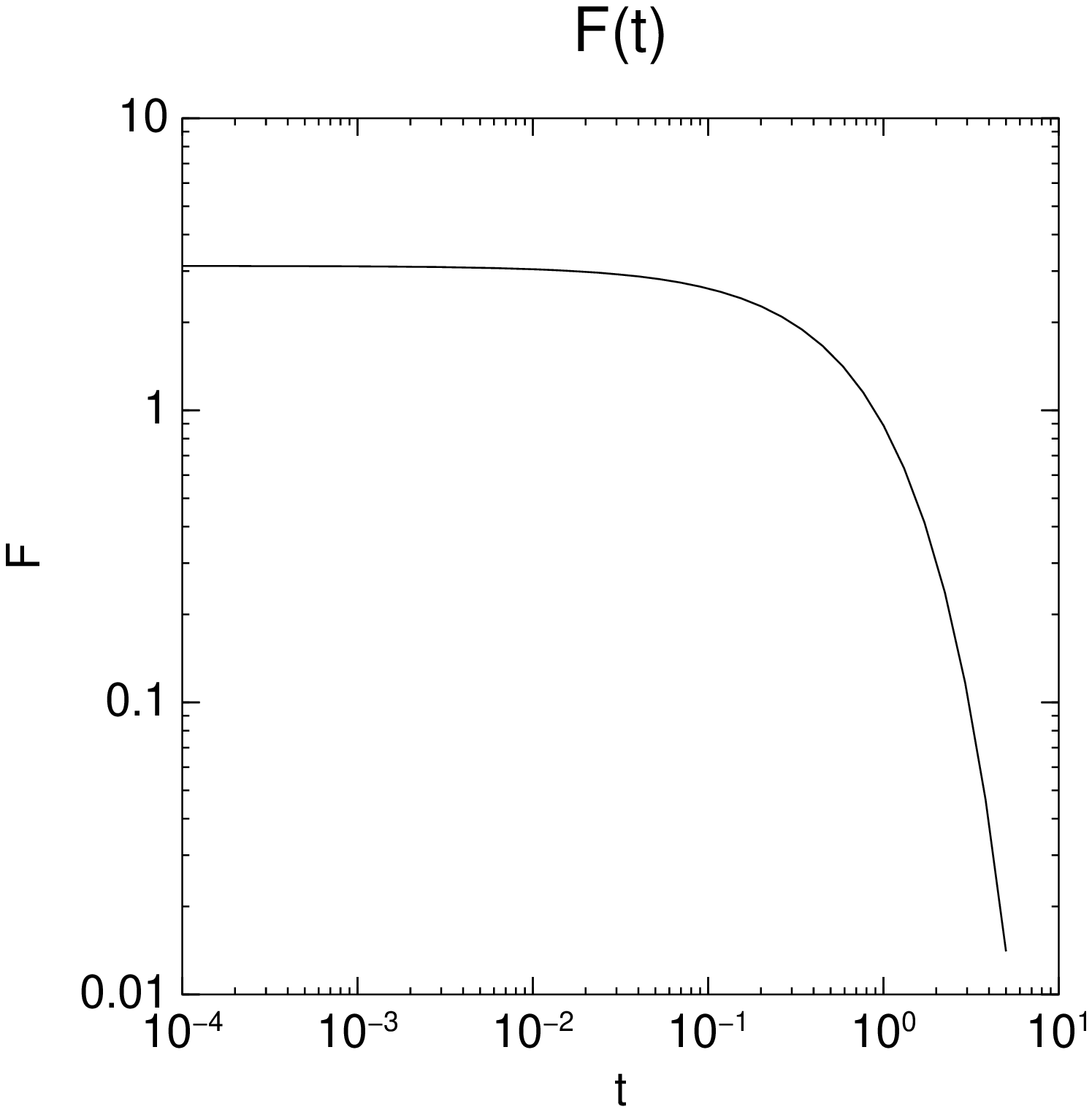}{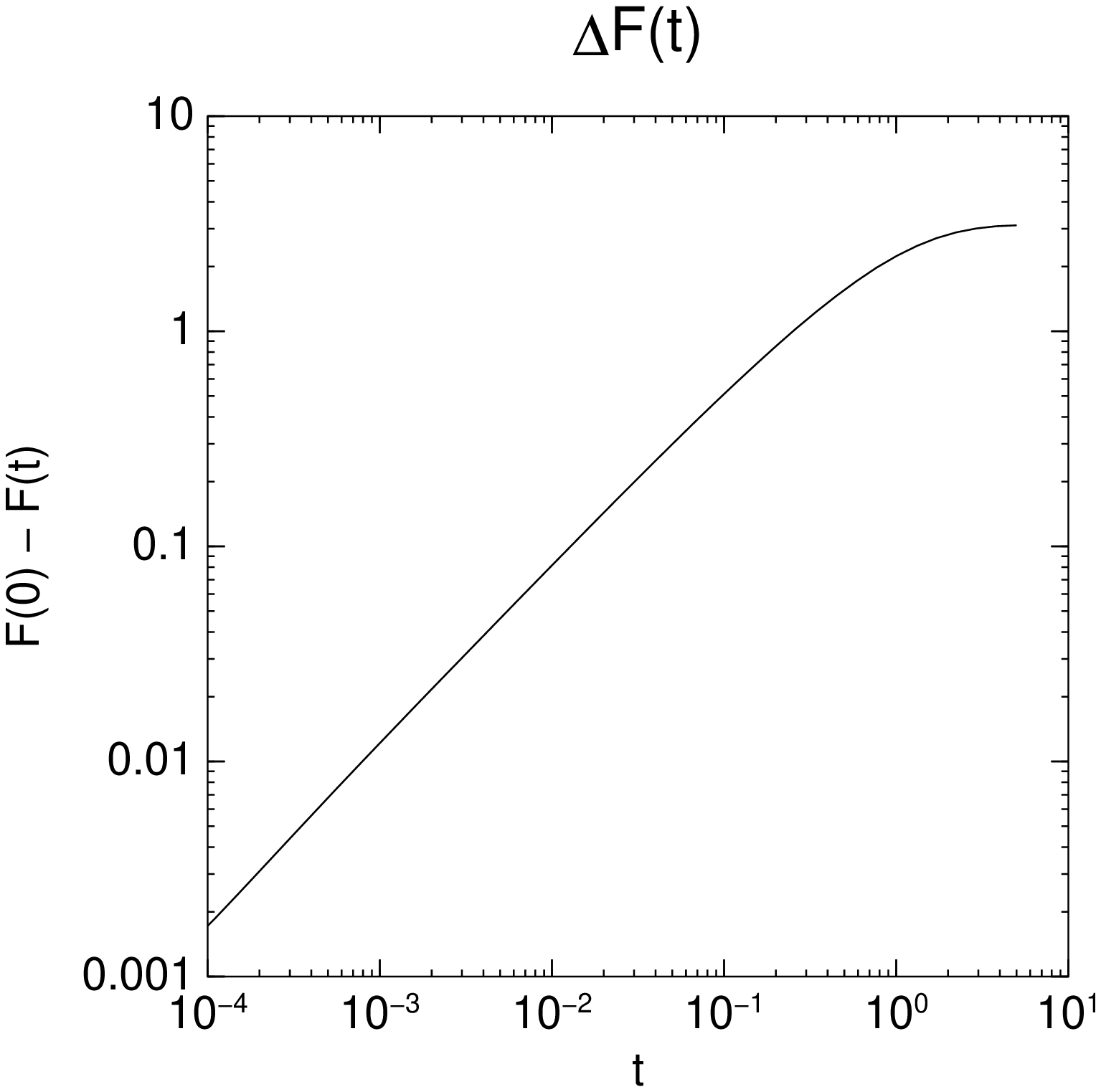}
  \caption{Synchrotron response function.  For $t < 1$ the function is roughly
    constant and $F(0) - F(t)$ is a power law with index $5/6$, while
    it exponentially decays at $t>1$.  This produces the synchrotron
    spectral indices of $-2/3$ at low energies and $-3/2$ at moderate
    energies, and the exponential tail at high energy.}
  \label{fig:f}
\end{figure*}

We can understand the synchrotron spectrum by examining its asymptotic
limits.  Figure \ref{fig:f} shows the function $F(t)$ and its
deviation from its zero-point value.  The function is nearly constant for
$t<1$, with $F(0)-F(t) \propto t^{5/6}$, and exponentially decreases
for $t<1$.  This divides the synchrotron response into three regimes.
If $\epsilon < \epsilon_i / (1+a^2) \lnlambda$, both $z_i$ and $z_f$
are less than one, and $N_\epsilon(\epsilon_i) \propto
\epsilon^{-2/3}$; this is the usual synchrotron spectral index.  If
$\epsilon_i / (1+a^2) \lnlambda < \epsilon < \epsilon_i / \lnlambda$,
$z_i < 1$ but $z_f > 1$, and $N_\epsilon(\epsilon_i) \propto
\epsilon^{-3/2}$; this reflects the decline of the particle Lorentz
factor.  If $\epsilon > \epsilon_i / \lnlambda$, both $z_i$ and $z_f$
are greater than 1, and $N_\epsilon(\epsilon_i)$ decays exponentially.

\subsection{Integral equation}

Since particles are assumed to be beamed along the field lines, the
specific number intensity at any point is
\begin{equation}
  I_\gamma(r, \psi, \epsilon) = \rho(r_e) Q_\gamma(r_e, \epsilon)
   e^{-\tau(r_e, r, \epsilon)}
  \label{eq:I_gamma}
\end{equation}
where $r_e$ is a function of $r$ and $\psi$ via equation
(\ref{eq:psi_of_r}), $Q_\gamma$ is the volumetric photon emission
rate, $\rho$ is the radius of curvature of the field line, and
$\tau(r_e, r, \epsilon)$ is the optical depth attained by
a photon of energy $\epsilon$ propagating from $r_e$ to $r$.  The
radius of curvature enters since $\rho d\psi$ is the path length
where the field points in direction $d\psi$.

The local volumetric rate of pair production events is then
\begin{equation}
  Q_{pp}(r, \epsilon) = \int_0^\infty d\psi \alpha_B(r, \psi, \epsilon)
    I(r, \psi, \epsilon).
\end{equation}
Using the synchrotron response, equation (\ref{eq:synch_n}), the
synchrotron emission rate is
\begin{equation}
  Q_{\gamma, syn}(r, \epsilon) = \int_0^\infty d\epsilon_i
    N_\epsilon(\epsilon_i) \int_0^\infty d\psi \alpha_B(r, \psi, \epsilon_i)
    I(r, \psi, \epsilon_i).
\end{equation}

Using equation (\ref{eq:I_gamma}) for $I_\gamma$, $d\tau = \alpha_B
\rho d\psi$, and the sharp-peaked nature of the absorption, we can
exactly evaluate the $\psi$ integral, yielding
\begin{equation}
  Q_{\gamma, syn}(r, \epsilon) = \int_0^\infty d\epsilon_i
    N_\epsilon(\epsilon_i) (1 - e^{-\tau_{max}})
    Q_\gamma(\bar{r}_e, \epsilon_i)
    \label{eq:basic_q_eq}
\end{equation}
where $\bar{r}_e = r_e(r, \bar{\psi})$, $\bar{\psi}$ is the peak of
the $\psi$ integral, and $\tau_{max}$ is the optical depth attained by
a photon of energy $\epsilon$ propagating from the stellar surface to
$r$.

Since, by equation (\ref{eq:dr_a}), a photon emitted at $r_e$ travels
at most $0.25 r_e$ before pair producing, we can treat the cascade as
if it occurred on-the-spot (OTS).  Photons with energies well above
$\epsilon_{min}$ pair-produce in a correspondingly shorter distance,
so the photons near $\epsilon_{min}$ set the spatial expanse of the
cascade.  This yields an integral equation for the synchrotron
spectrum,
\begin{eqnarray}
  Q_{\gamma, syn}(\epsilon) & = & \int_0^\infty d\epsilon_i
    (1 - e^{-\tau_{\infty}(\epsilon_i)}) \times \nonumber \\
    & & \frac{1}{\epsilon_i} K(\frac{\epsilon}{\epsilon_i})
    (Q_{\gamma, src}(\epsilon_i) + Q_{\gamma, syn}(\epsilon_i))
    \label{eq:integral_eq}
\end{eqnarray}
where we have replaced $\tau_{max}$ with $\tau_{\infty}$ and where
\begin{eqnarray}
  K(\frac{\epsilon}{\epsilon_i}) & = &
    \frac{3\sqrt{3}}{8 \pi} \sqrt{\lnlambda}
    \left(\frac{\epsilon}{\epsilon_i}\right)^{-3/2} \times \nonumber \\
    & & \left(F(\lnlambda \frac{\epsilon}{\epsilon_i}) -
      F(\phi \lnlambda \frac{\epsilon}{\epsilon_i}) \right)
\end{eqnarray}
where $\phi \equiv (1+a^2)$.

This can be easily solved by matrix operator methods.  If
$Q_{\gamma,syn}$ and $Q_{\gamma,src}$ are represented by vectors
$\bar{Q}_{syn}$ and $\bar{Q}_{src}$, and the integral operator above
by a matrix $\hat{K}$, the solution is simply $\bar{Q}_{syn} = (1 -
\hat{K})^{-1} \hat{K} \bar{Q}_{src}$.

\begin{figure*}
  \plotone{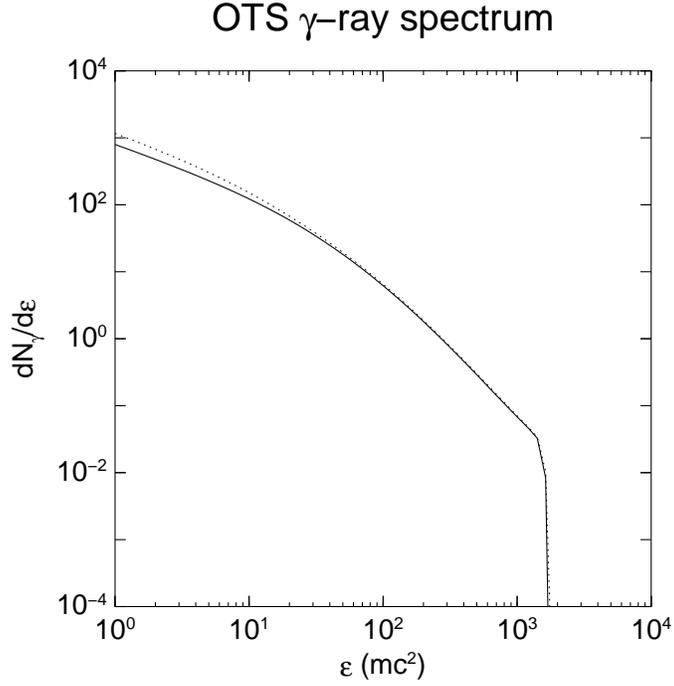}
  \caption{The $\gamma$-ray spectrum produced by
    a single input photon, computed by the OTS model (solid line) and the
    full numerical model (dotted line), evaluated for $P = 0.1$ s,
    $B = 10^{12}$ Gauss, and
    $\epsilon = 10^{3} \epsilon_a = \sci{6.5}{5} \, mc^2$.}
  \label{fig:ots}
\end{figure*}

This gives an excellent approximation to the final $\gamma$-ray
spectrum, and a close approximation to the final pair spectrum, as
shown in Figure \ref{fig:ots}.  The numerically calculated pair
spectrum extends to lower energies than the on-the-spot (OTS)
spectrum, due to the change in magnetic field over the course of the
cascade.

\begin{figure*}
  \plotone{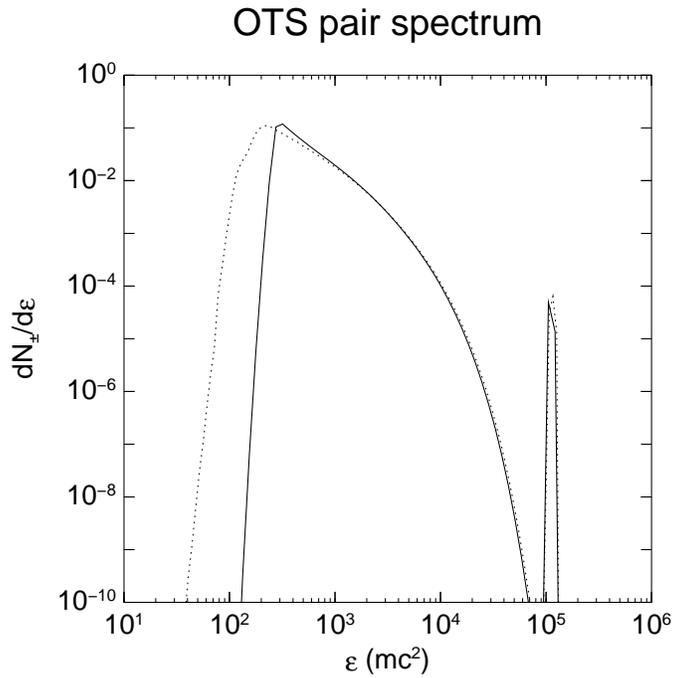}
  \caption{The pair spectrum produced by
    a single input photon, computed by the OTS model ({\it solid line}) and the
    full numerical model ({\it dotted line}), evaluated for $P = 0.1$ s,
    $B = 10^{12}$ Gauss, and
    $\epsilon = 10^{3} \epsilon_a = \sci{6.5}{5} \, mc^2$.}
\end{figure*}

\subsection{Moment equations}

From the integral equation for the synchrotron photon distribution, we
find several useful relations connecting the moments of the photon
spectrum.  Using equation (\ref{eq:integral_eq}) to calculate
successive generations of synchrotron photons, we find
\begin{equation}
  Q_{\gamma}^{(i+1)}(\epsilon) = \int_{\epsilon_{min}}^{\infty} d\epsilon_i \,
    \frac{1}{\epsilon_i} K(\frac{\epsilon}{\epsilon_i})
    Q_{\gamma}^{(i)}(\epsilon_i)
\end{equation}
where $Q_{\gamma}^{(i)}$ is the $i^{th}$-generation $\gamma$-ray
spectrum and where we have replaced $1-\exp(-\tau_\infty)$ with a step
function at $\epsilon_{min}$.

Multiplying by $\epsilon^n$ and integrating the above expression over
all energies gives
\begin{equation}
  Q_n^{(i+1)} = R_n^{(i)} K_n,
\end{equation}
where
\begin{eqnarray}
  Q_n^{(i)} & \equiv & \int_0^\infty d\epsilon \,
                       \epsilon^n Q_{\gamma}^{(i)}(\epsilon) \\
  R_n^{(i)} & \equiv & \int_{\epsilon_{min}}^\infty
                       d\epsilon \, \epsilon^n Q_{\gamma}^{(i)}(\epsilon) \\
  K_n       & \equiv & \int_0^\infty dx \, x^n K(x).
\end{eqnarray}
$R_n$ is the ``reduced moment'' of the spectrum, neglecting energies
below $\epsilon_{min}$.

Since $K$ is known, the $K_n$ are known.  Given a known source
function, $R_n^{(0)}$ may be calculated.  The moments of successive
generations may then be calculated by assuming that $R_n \approx Q_n$
so that $Q_n^{(i)} = K_n^i Q_n^{(0)}$.  The limited number of
generations truncates the sum of successive moments.

Integrating the kernel, we find that the first two moments are
\begin{eqnarray}
  K_0 & = & \frac{15 \sqrt{3}}{8} \lnlambda (\phi^{1/2} - 1) \\
  K_1 & = & 1 - \phi^{-1/2} \label{eq:k_1}
\end{eqnarray}
In general, the $n^{th}$ moment is
\begin{eqnarray}
  K_n & = & \frac{9\sqrt{3}}{8 \pi} \frac{2^n \lnlambda^{1-n}}{2n^2 + n - 1}
  \left(1 - \phi^{\frac{1}{2} - j}\right) \times \nonumber \\
  & &
  \Gamma\left(\frac{n}{2} + \frac{1}{6}\right)
  \Gamma\left(\frac{n}{2} + \frac{11}{6}\right)
\end{eqnarray}
From $K_0$ and $K_1$ respectively, we can see that the total number of
photons steadily increases while almost all of the energy remains
in the photon spectrum, provided that $\phi = (1+a^2)$ is large.  The
fraction of energy left in the pair spectrum is $1-K_1$ =
$\phi^{-1/2}$, which becomes significant for values of $\phi$ near 1,
which occurs at magnetic fields greater than \sci{3}{12} Gauss.

\subsection{Generations}

A cascade of pair production then divides into several successive
generations of photons converting into pairs which emit more photons.
Following the work of \citet{tademaru73} and later
papers \citep{lu94, wei97, zhang00}, we find, in our notation, that
the characteristic synchrotron energies of successive generations are
related via
\begin{equation}
  \epsilon_{j+1} = \frac{\epsilon_j}{\lnlambda}
  \label{eq:e_generations}
\end{equation}
while the particles produced by generation $j$ have final Lorentz factors
\begin{equation}
  \gamma_{fj} = \frac{\epsilon_j}{2 \sqrt{1 + a^2}} \approx
    \frac{\epsilon_j}{2 a}
  \label{eq:gamma_generations}
\end{equation}
where the approximate form assumes $a$ is well greater than 1, which
is true provided that $B < \sci{3}{12}$ Gauss.  These correspond
exactly to the expressions cited by \citet{zhang00}, given the
equivalence $\lnlambda = 4/3 \chi$.

From this, it is clear that for each absorbed photon, a fraction of
the energy,
\begin{equation}
  f_{\gamma} = 1 - \frac{1}{\sqrt{1 + a^2}} = K_1
  \label{eq:f_gamma}
\end{equation}
is immediately re-emitted in lower-energy photons, with the remainder
left in the electron-positron pair.  As discussed by \citet{zhang00},
much of the the energy in the pairs is later radiated at
energies near $2 \epsilon_b \gamma_f$ via resonant inverse Compton
scattering, but this process generally does not create further pairs.

These approximations effectively collapse the entire synchrotron
emission of a created pair into a pulse of photons at the peak energy
of the synchrotron spectrum.  By using the moments of the photon
spectra, we can
derive similar approximations, but with wider applicability.

At each generation, the number of photons increases by a factor of
$K_0$, while the total energy in the photons decreases by a factor of
$K_1$, so the average energy of the $j^{th}$ generation is
\begin{equation}
  \bar{\epsilon}_{j} = \bar{\epsilon}_0 \left(\frac{K_1}{K_0}\right)^{j}
  \label{eq:energy_j}
\end{equation}
where $\bar{\epsilon_0}$ is the average energy of the input spectrum.
The total number of photons in the $j^{th}$ generation is
\begin{equation}
  N_{j} = N_0 K_0^j
  \label{eq:multiplicity_j}
\end{equation}
where $N_0$ is the initial number of photons.  Provided that
$\bar{\epsilon}_j > \epsilon_a$, most of the photons will convert into
pairs.  If we
truncate the generations when the average energy $\bar{\epsilon}_{j}$
reaches $\epsilon_a$,
we find a power-law relationship between the initial energy and the final
pair multiplicity,
\begin{equation}
  M_{tot}(\epsilon_0) \propto \left(\frac{\epsilon_0}{\epsilon_a}\right)^\nu
  \label{eq:mult_power_law}
\end{equation}
where the power-law exponent is given by
\begin{equation}
  \nu = \frac{\ln K_0}{\ln K_0 - \ln K_1}.
  \label{eq:nu}
\end{equation}
The constant of variation depends on the behavior of the lowest-energy
photons near $\epsilon_{min}$ and will be discussed in Section
\ref{sec:one_photon_pairs}.

We can apply similar arguments to the {\em spatial} extent of the
cascade.  From equation (\ref{eq:dr_a}), the photons of generation $j$,
with mean energy $\bar{\epsilon}_j$, must propagate a distance
$\Delta s = 0.25 (\epsilon_a / \epsilon_j) R_*$ before pair-producing.
Relating the multiplicity of generation $j$, equation
(\ref{eq:multiplicity_j}), to this propagation distance gives a power-law
dependence between the multiplicity and the distance,
\begin{equation}
  M(s) \propto s^\nu
  \label{eq:mult_s_power_law}
\end{equation}
where the power-law exponent, $\nu$, is identical to that of the
energy-multiplicity relation, equation (\ref{eq:nu}).

\section{Numerical method}

Since the OTS method cannot spatially resolve the pair cascade, we turn
to a numerical calculation.  Equation (\ref{eq:basic_q_eq}) provides
the fundamental structure for the calculation.  It gives the synchrotron
spectrum injected into the magnetosphere at any point, as a function
of the photons absorbed at that point.

To calculate the resultant spectrum, we simply evaluate the photon spectrum
$Q(\epsilon)$ at logarithmically spaced points in $\epsilon$.  The relation
between the synchrotron photons output and the absorbed spectrum at a given
radius is then a simple matrix multiplication by $N_\epsilon(\epsilon')$.
Due to the variation of magnetic field with altitude, this matrix will also
depend on the altitude.

We then discretize the radial steps on a variable scale, with either
purely logarithmic steps, in cases where we are primarily concerned with
the $\gamma$-ray output and the location of the PFF, or a combination of
logarithmic steps, up to a polar-cap radius, and linear steps beyond, to
calculate a smooth pair spectrum.  Since the final pair energies depend on
the local magnetic field, the pair spectrum is more sensitive to the spatial
variations in the field, requiring this scheme.

To begin the calculation, at each radial point we inject an initial
spectrum.  These spectra are either a delta-function source at the
lowest radius, for measuring the single-photon response, or the
expected spectra emitted by a single beam particle as it passes
through each radial bin, for a full physical cascade.  In the second case,
the spectrum is adjusted to prevent binning effects from affecting the
total power injected.

Using the expression $\tau_\infty$, equation (\ref{eq:tau_infty}), we
find which photons are absorbed within the magnetosphere, and which
escape.  The escaping photons are simply accumulated.  The absorbed
photons are assumed to be absorbed at $\psi(\epsilon) =
\psi_a(\epsilon)$, or at a radius of $r' = r + \rho(r) \psi_a$.  Each
photon energy from each radial bin is absorbed at a different height;
these final heights are then simply re-binned into the grid.

For each absorbed photon, equation (\ref{eq:synch_n}) is used to
determine the synchrotron spectrum injected at the absorption point.
This process is then iterated, to find the secondaries produced by
those synchrotron photons, and so on, until the energy remaining in
synchrotron photons is negligible.

These calculations are streamlined by pre-calculating the synchrotron
response matrices, $N_\epsilon(\epsilon_i; r)$, and the total
absorption, $\tau_\infty(\epsilon; r)$, at each altitude.  In
addition, the radial absorption matrix, $r'(r; \epsilon)$, is computed
for each energy, reducing each step of the iteration to a series of
matrix multiplications.

\begin{figure*}
  \plotone{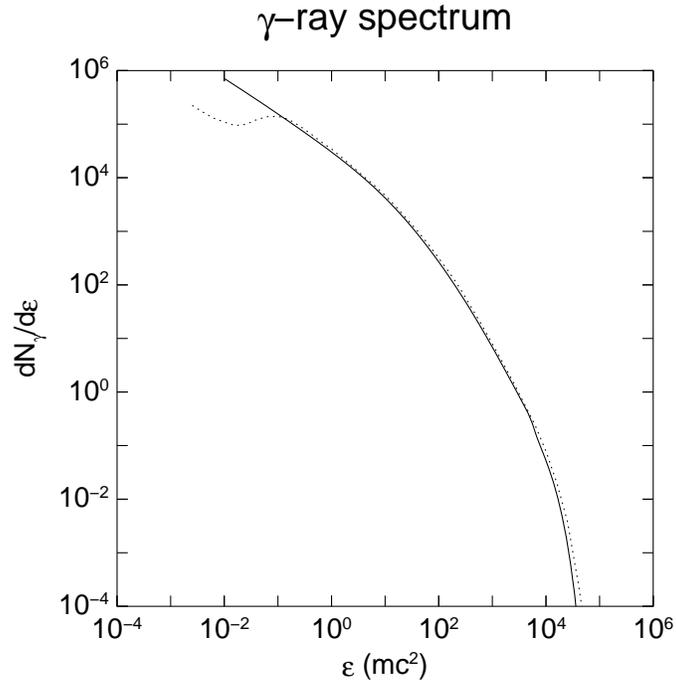}
  \caption{Predicted final $\gamma$-ray spectrum resulting from injecting
    a single particle with Lorentz factor $\gamma = \sci{6}{7}$ at the
    surface of a pulsar with a surface magnetic field of $10^{12}$ Gauss and
    field line radius of curvature of $10^8$ cm.  The solid line represents
    the model of this paper, while the dotted line represents the results of
    the \citet{daugherty82} Monte Carlo model.}
  \label{fig:harding_gamma}
\end{figure*}

\begin{figure*}
  \plotone{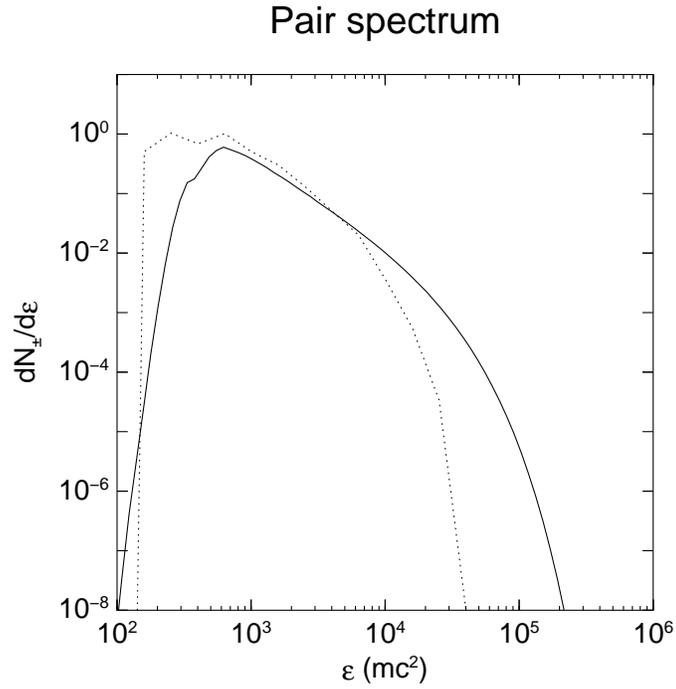}
  \caption{Predicted final pair spectrum resulting from injecting
    a single particle with Lorentz factor $\gamma = \sci{6}{7}$ at the
    surface of a pulsar with a surface magnetic field of $10^{12}$ Gauss and
    field line radius of curvature of $10^8$ cm.  The solid line represents
    the model of this paper, while the dotted line represents the results of
    the \citet{daugherty82} Monte Carlo model.}
  \label{fig:harding_pairs}
\end{figure*}

Figures \ref{fig:harding_gamma} and \ref{fig:harding_pairs} show,
respectively, the final $\gamma$-ray and pair spectra for both this
model and the Monte Carlo model of \citet{daugherty82}.
The predicted $\gamma$-ray spectra match extremely well, while the
pair spectra mismatch at low energies, due to the low resolution of the
Monte Carlo method in that range, and at high energies, due to the
approximation in this model of treating $\lnlambda$ as a constant,
rather than allowing it to vary with photon energy.

\section{Single-photon results}
\label{sec:one_photon}

To understand the cascade process itself, we first examine the response
to a single photon of injected into the magnetosphere at the surface
of the star.  Physically, the resultant cascade depends on the energy of
the input photon, the local magnetic field strength, and the local
radius of curvature of the field lines.

\subsection{Gamma-ray spectra}
\label{sec:one_photon_gamma}

From the discussion of the optical depth, we know that only photons
emitted at radius $r$ with energies less than $\epsilon_{min}(r)$ will
escape the magnetosphere.  Due to the decline of the magnetic field
with $r$, $\epsilon_{min}$ steadily increases with altitude.

However, the cascade produced by a single injected high-energy photon
remains localized.  If there are multiple generations of pair
production, the photon energy must be greater than $\lnlambda
\epsilon_a$, from equation (\ref{eq:e_generations}), which
pair-produces in $\Delta r < r_e / 4 \lnlambda$.  Since $\lnlambda$ is
on the order of 20, this distance is short compared to $r_e$, and the
magnetic field remains effectively unchanged.  Hence, we may treat the
cascade process as if it occurred on-the-spot, and the energy cut-off
remains at $\epsilon_{min}(r_e)$.

The typical synchrotron response has a power-law exponent $\nu =
-3/2$, extending from energy $\epsilon_i/\lnlambda$ to
$\epsilon_i/(1+a^2)\lnlambda$, and a power-law exponent of $\nu =
-2/3$, the typical value for synchrotron radiation, for lower
energies.  After processing through multiple generations, this
translates to an output spectrum with $\nu = -2/3$ up to
$\epsilon_{min}/(1+a^2)\lnlambda$ and $\nu = -3/2$ from there to
$\epsilon_{min}$.

\begin{figure*}
    \plotone{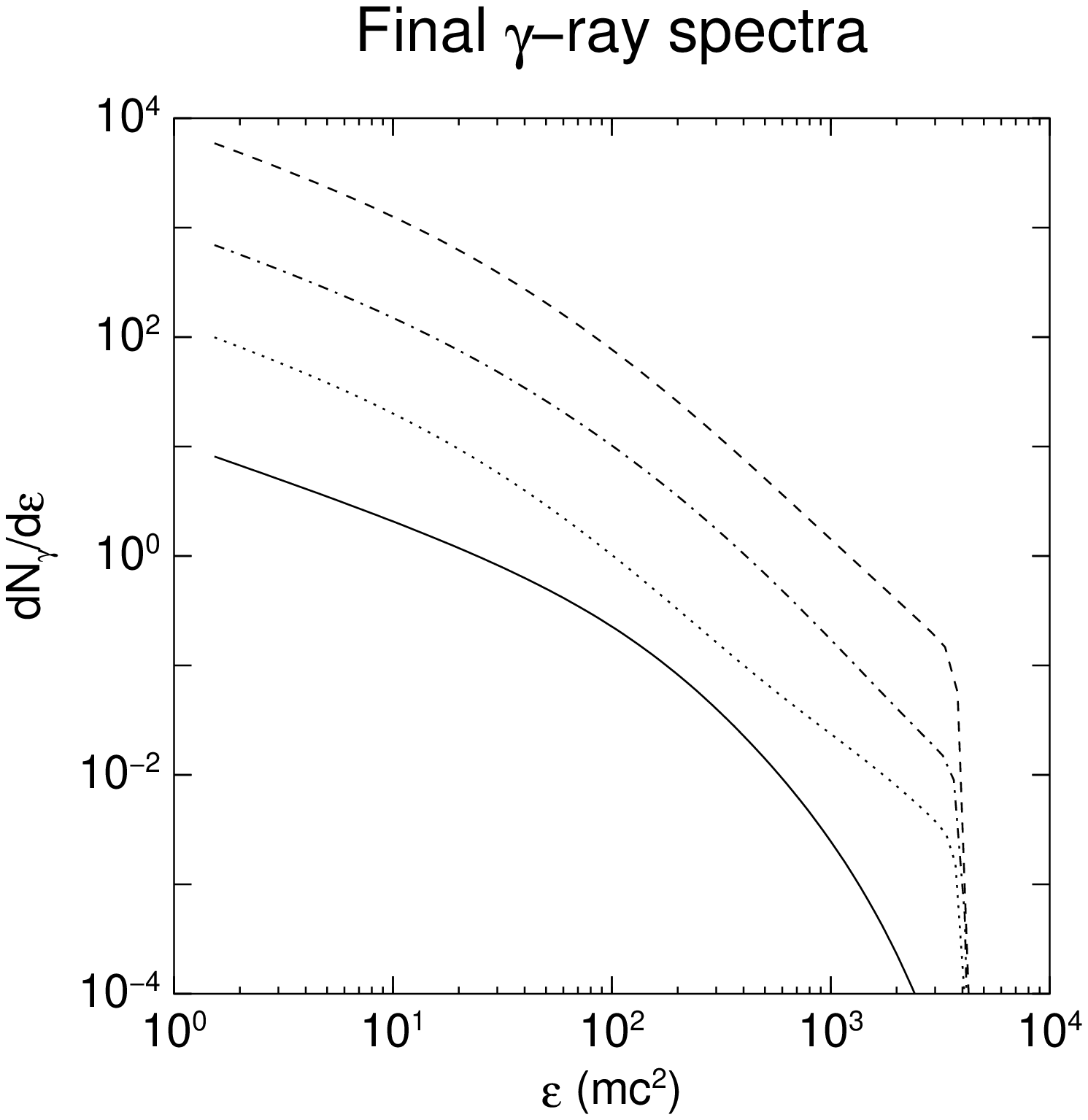}
    \caption{Final $\gamma$-ray spectrum produced by injecting a single photon
    at the stellar surface, evaluated for $B=10^{12}$ Gauss and
    $\Phi_{cap} = 10^{14}$ V.  The input photon energies used are $10$,
    $10^2$, $10^3$, and $10^4$ times $\epsilon_a$ with higher input energies
    correspond to higher amplitudes.  In this case, $\epsilon_a = 1530\,mc^2$
    and $\epsilon_{min} = (64/27) \epsilon_a = 3627\,mc^2$.  This clearly
    shows the sharp transition from free to absorbed photons at
    $\epsilon_{min}$.
    }
    \label{fig:gamma_for_various_es}
\end{figure*}

Figure \ref{fig:gamma_for_various_es} shows the gamma-ray spectra
produced by various input energies.  This demonstrates the expected
cut-off at $\epsilon_{min}$, as well as the transition from a power-law
exponent of $-3/2$ to $-2/3$ at $\epsilon_{min}/(1+a^2)\lnlambda$.
The shape of the spectrum is different at low energies, due to the
effects of the exponential tails of the opacity and synchrotron
emission, but it quickly converges to the asymptotic form.

\subsection{Pair multiplicity}
\label{sec:one_photon_pairs}

Since the total photon energy at each pair-production event is roughly 
conserved, cascades in low- and mid-field pulsars effectively convert 
high-energy photons into photons with energy $\epsilon_{min}$. At the 
penultimate generation of an extended cascade, however, a significant 
fraction of the synchrotron photons have $\epsilon < \epsilon_{min}$ and 
escape. From the $-3/2$ synchrotron power law and using the expected 
multiplicity power-law dependence on energy, equation
(\ref{eq:mult_power_law}), we expect a final multiplicity of
\begin{equation}
  M_{tot} = 1 + \frac{1}{\sqrt{\lnlambda}}
      \left(\frac{\epsilon}{\epsilon_{min}}\right)^\nu,
   \label{eq:m_tot}
\end{equation}
with $\nu$ as given in equation (\ref{eq:nu}).

For pulsars with moderate and low fields, $B \lesssim \sci{3}{12}$,
the quantity $a$ is large, a negligible fraction of the energy of
each absorbed photon is left in the generated pair, and $\nu \approx 1$,
yielding a simple linear relation between input photon energy and
final pair multiplicity.

In the opposite regime, very high fields, $a$ is small, so the
generated pair retains almost all of the initial photon energy, and
there is little or no cascade; high-energy photons simply transmute
into high-$\gamma$ particles.

\begin{figure*}
  \plotone{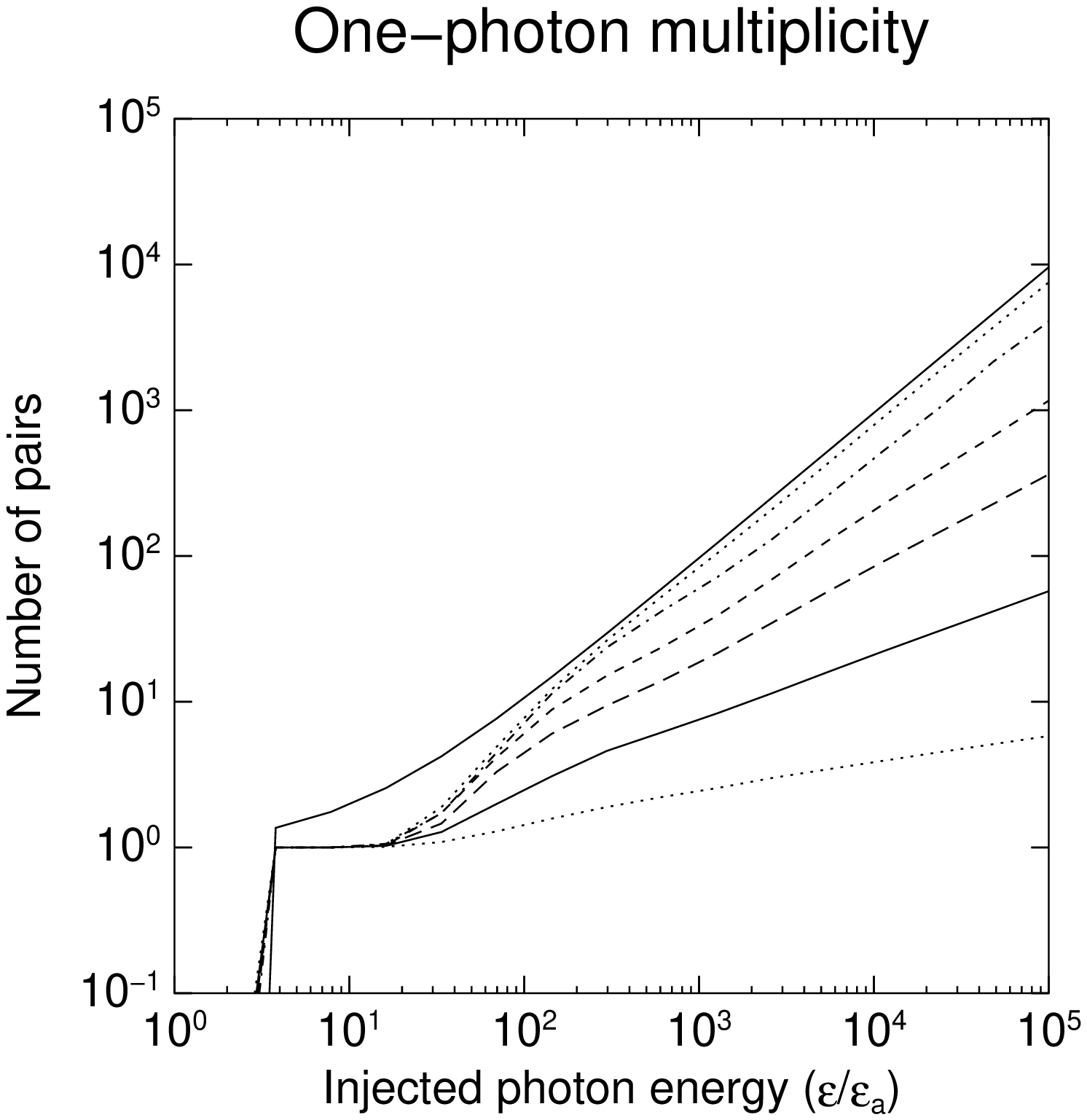}
  \caption{Final pair multiplicity produced by injecting a single
    photon of varying energy at the stellar surface.  The topmost solid line
    shows the theoretical prediction for $B=10^{11}$ Gauss, while the other
    lines represent the numerical results for
    different magnetic field strengths: \sci{1}{11},
    \sci{1}{12}, \sci{2}{12}, \sci{3}{12}, \sci{5}{12}, and \sci{1}{13}
    Gauss, respectively.  The decrease in the power-law slope of the response
    with increasing magnetic field is clearly visible.}
  \label{fig:ratio_mult}
\end{figure*}

In Figure \ref{fig:ratio_mult}, we plot the numerically calculated
energy dependence of the total multiplicity along with the theoretical
results.  The simplest multiplicity
formula, equation (\ref{eq:m_tot}) with the exponent $\nu$ set to 1,
works well for magnetic
fields below $10^{12}$ Gauss, but over-estimates the multiplicity for
higher fields, due to the increase in the amount of energy left in the
particles in the higher-field cascades.  For all magnetic fields,
the spectrum has three well-defined regions.  If
$\epsilon < \epsilon_{min}$, the initial photon is below threshold,
and no pairs are produced.   If $\epsilon_{min} < \epsilon < \lnlambda 
\epsilon_{min}$, the initial photon will pair produce, but all of the
secondary synchrotron photons have energies below $\epsilon_{min}$ and
create no further pairs.  Finally, for
$\epsilon > \lnlambda \epsilon_{min}$, both the initial photon and the
secondary synchrotron photons may create pairs, producing a smoothly
increasing multiplicity.

\begin{figure*}
  \plotone{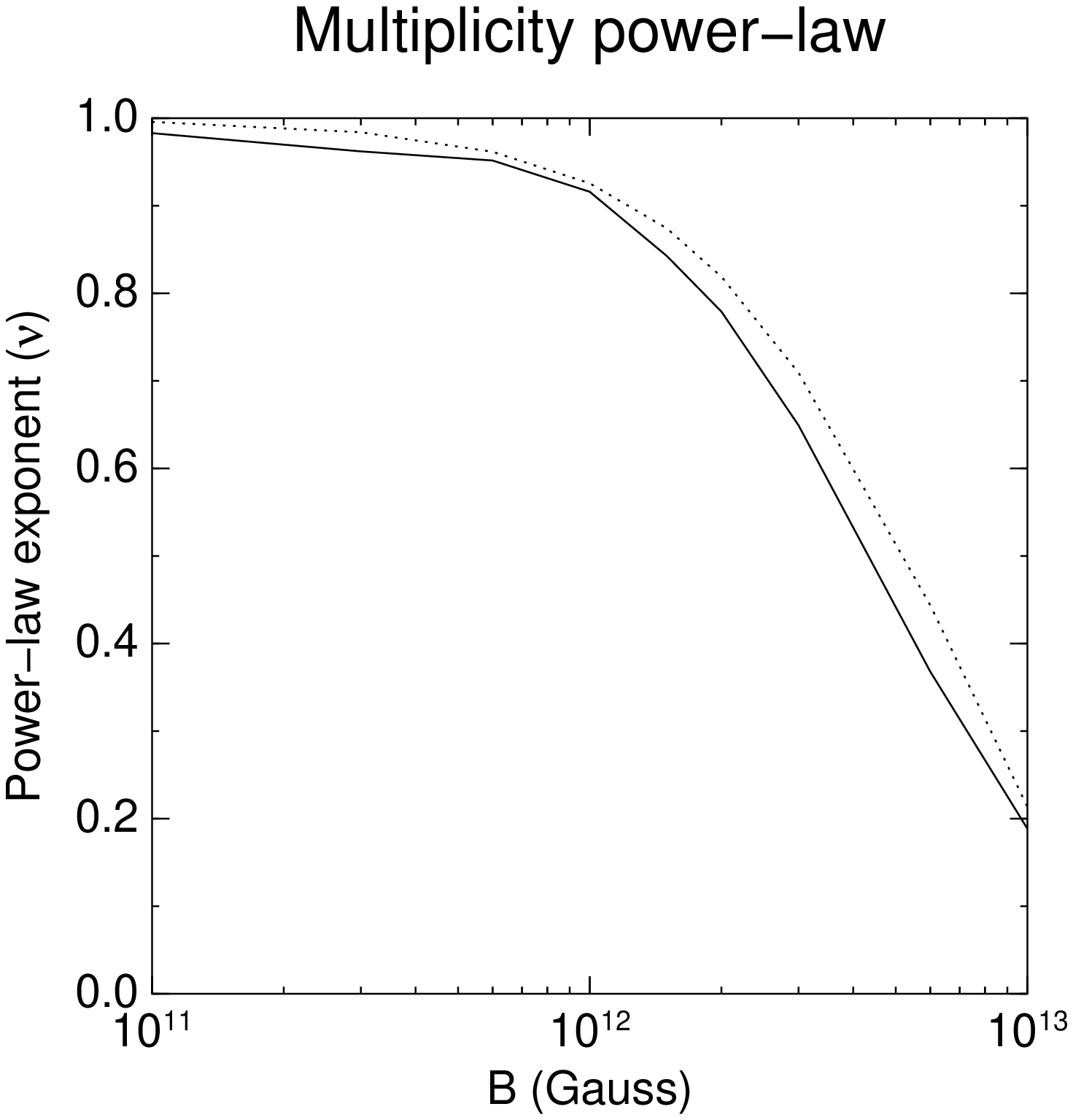}
  \caption{Final pair multiplicity power-law exponent, $\nu$, where
    $M_{tot}(\epsilon) \propto \epsilon^\nu$.
    The solid line shows the computed value, while the dotted shows the
    theoretical prediction.}
  \label{fig:power_law_bs}
\end{figure*}

Figure \ref{fig:power_law_bs}
shows this change in the power-law exponent with magnetic field, along
with the theoretical result, equation (\ref{eq:nu}).  The approximation
closely matches the numerical results over the entire dynamic range.

From equation (\ref{eq:mult_s_power_law}), we expect that the spatial
development of the cascade should be a  similar power-law in space.  To
find a closer approximation, we use an argument akin to that for the
final multiplicity.

Since the first several generations of a high-energy cascade occur in
the high-energy regime where equation (\ref{eq:dr_a}) applies, the
cascade develops as if $\epsilon_a$ were the minimum energy, only
deviating when the photon energy drops near $\epsilon_{min}$.  Hence,
a good approximation to the high-energy cascade is
\begin{eqnarray}
  M(\epsilon, s) & = & 1 + \frac{1}{\sqrt{\lnlambda}}
    \left(\frac{\epsilon}{\epsilon_a} \frac{4 (s-s_1)}{r_e}\right)^\nu,
    \nonumber \\ 
    & & s_1 < s < s_1 + \frac{27}{64} \frac{r_e}{4}
    \label{eq:mult_s_approx}
\end{eqnarray}
where $s$ is the propagation distance from the emission point,
$s_1$ is the point at which the initial photon pair produces,
$s_1 = 0.25 (\epsilon_a/\epsilon) r_e$, $\nu$
is given in equation (\ref{eq:nu}), and the
factor of $27/64$ follows from the truncation of the cascade at
$\epsilon_{min}$ rather than $\epsilon_a$.

\begin{figure*}
  \plotone{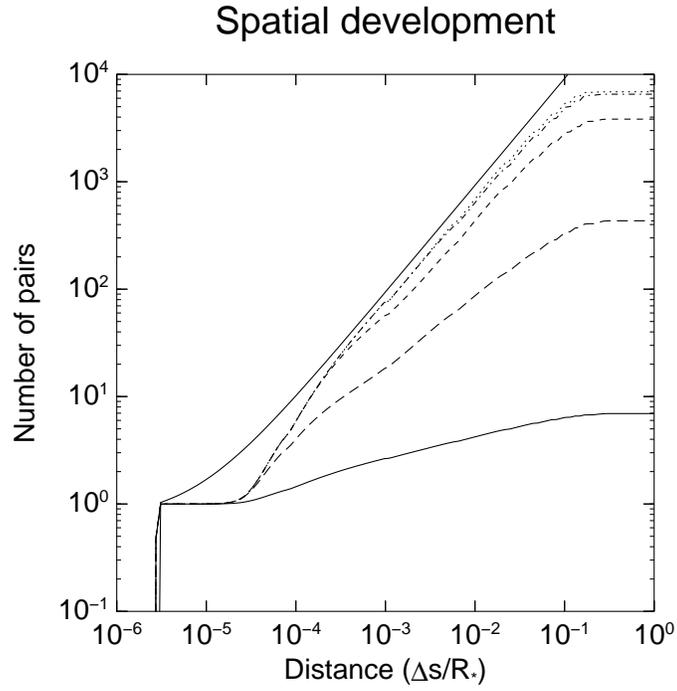}
  \caption{Cumulative pair production as a function of distance, for several
  values of the magnetic field B and $\Phi_{cap} = 10^{14}$ V. The
  topmost solid line shows the
  theoretical prediction for $B=10^{11}$ Gauss, while the other lines represent
  different magnetic field strengths: \sci{1}{11}, \sci{3}{11},
  \sci{1}{12}, \sci{3}{12}, and \sci{1}{13} Gauss, respectively.}
  \label{fig:spatial_cascade}
\end{figure*}

\begin{figure*}
  \plotone{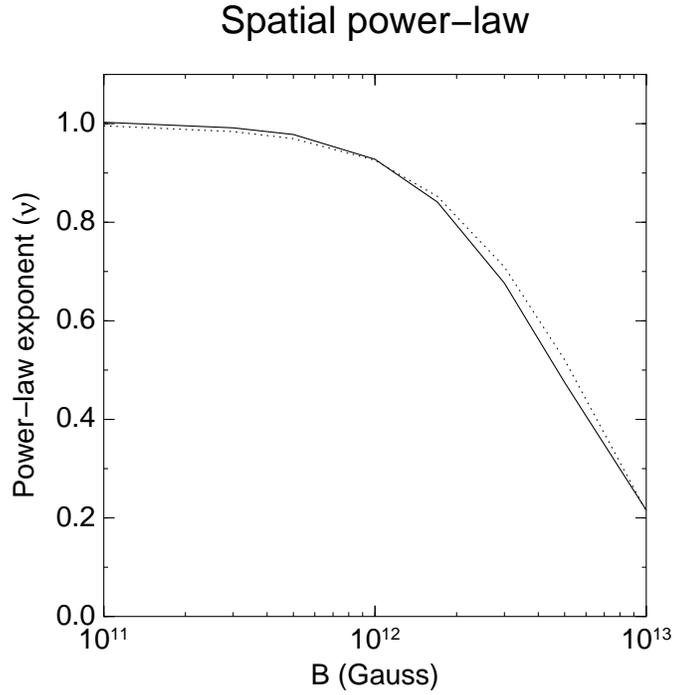}
  \caption{Variation of the spatial power-law exponent
  with B.  The solid line is the computed power-law exponent and the
  dotted line the theoretical prediction.}
  \label{fig:spatial_nu}
\end{figure*}

Examining the numerical results, shown in Figure
\ref{fig:spatial_cascade}, we found that, indeed, the pair production
rate follows this rule very well.  The total length of the cascade is
approximately $0.11\,R_*$, which is the pair production distance
expected for a photon of the minimum energy capable of pair-producing,
$64 \epsilon_a / 27$, according to the $\psi_a$ model.  The computed
power law, shown in Figure \ref{fig:spatial_nu} follows the predicted
values almost exactly.

\subsection{Pair spectra}

\begin{figure*}
    \plotone{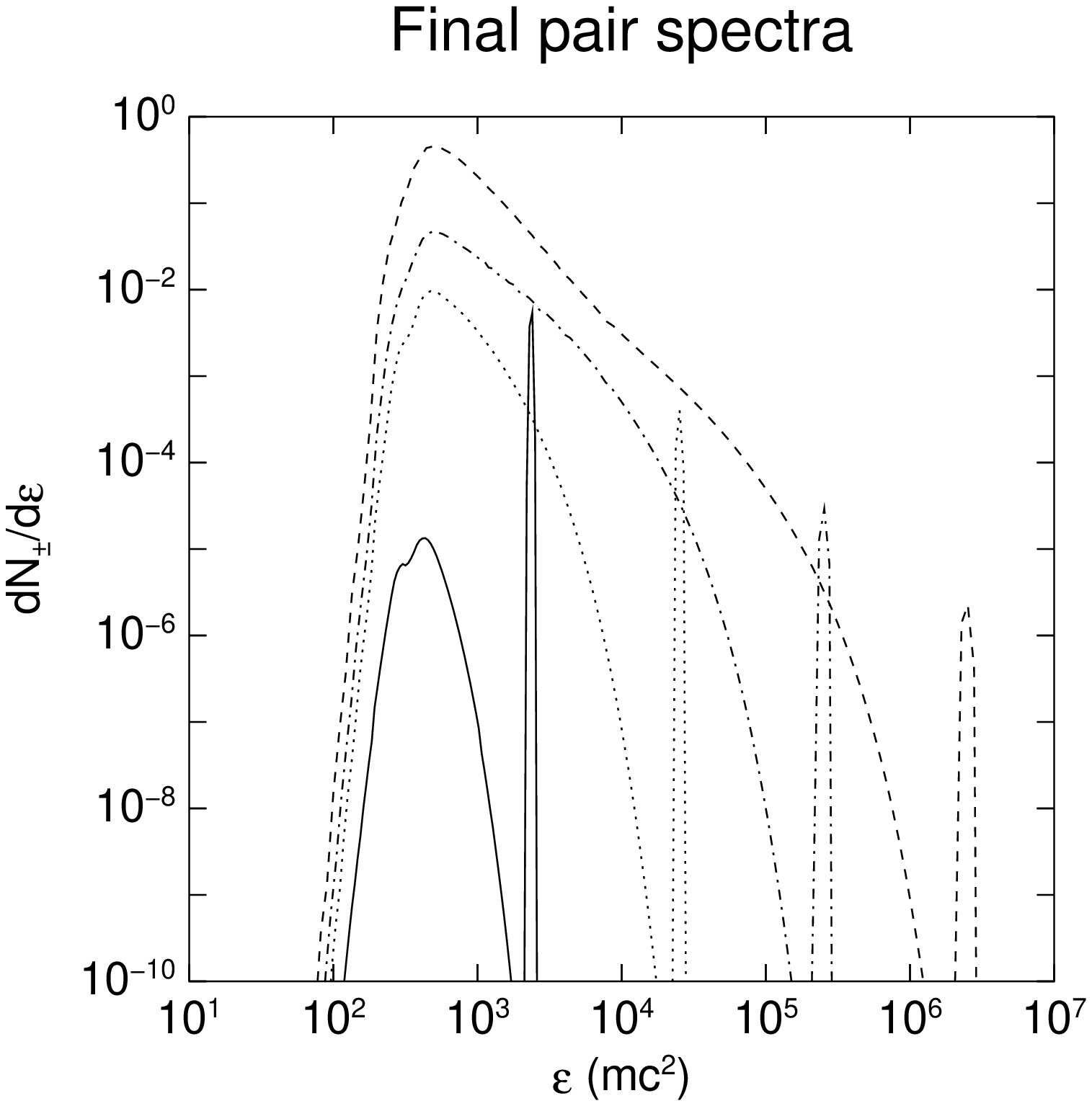}
    \caption{Final pair spectra produced by injecting a single photon at
    the stellar surface, evaluated at a $B=10^{12}$ Gauss and
    $\Phi_{cap} = 10^{14}$ V, for various input photon energies.  The input
    photon energies used are $10$,
    $10^2$, $10^3$, and $10^4$ times $\epsilon_a$ with higher input energies
    correspond to higher amplitudes. For these parameters,
    $\epsilon_a = 1530\,mc^2$.
    }
    \label{fig:pairs_for_various_es}
\end{figure*}

Figure \ref{fig:pairs_for_various_es} shows the variation in the
output pair spectrum, for different values of the input energy.  The
pair spectra are characterized by a sharp rise near $\epsilon_a/a$,
followed by a $\nu=-3/2$ power law up to $\epsilon_0/\lnlambda$, where
$\epsilon_0$ is the energy of the injected photon, followed by an
exponential decline.  The spike corresponding to the input photon is
also visible.

\begin{figure*}
    \plotone{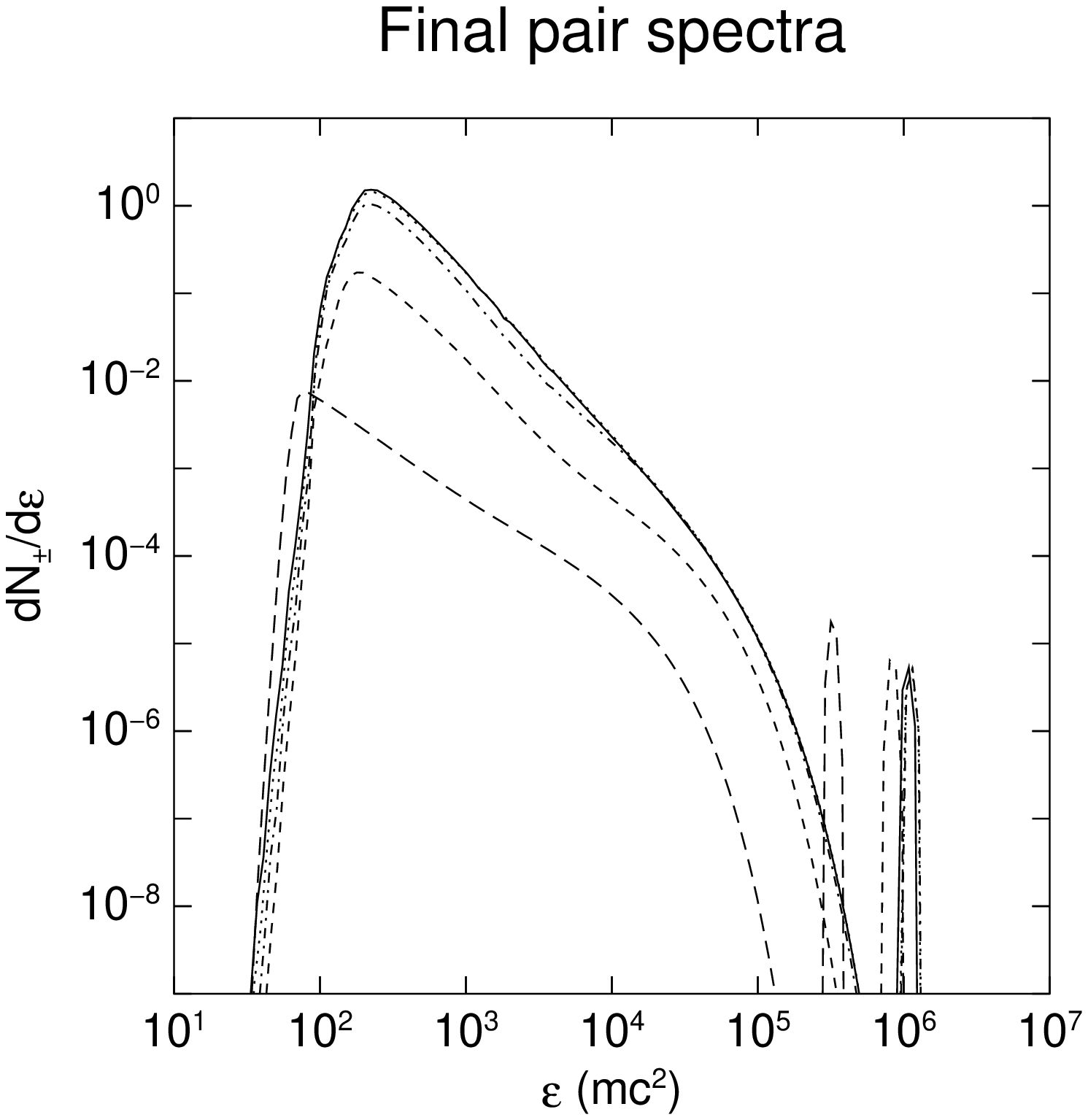}
    \caption{Final pair spectra produced by injecting a single photon at
    the stellar surface, evaluated at fixed voltage,
    $\Phi_{cap} = 10^{14}$ V,
    and fixed input energy ratio, $\epsilon = 10^4 \, \epsilon_a(B)$, for various
    values of the magnetic field.  The magnetic fields used are \sci{1}{11},
    \sci{3}{11}, \sci{1}{12}, \sci{3}{12}, and \sci{1}{13} Gauss, from highest
    peak amplitude to lowest.}
    \label{fig:pairs_for_various_bs}
\end{figure*}

Figure \ref{fig:pairs_for_various_bs} shows the variation in the
output pair spectrum, for a fixed ratio of input photon energy to
$\epsilon_a$, $\epsilon_0/\epsilon_a = 10^4$, for different values of
the magnetic field.  Here, we notice the similarity of the pair
spectra at different field strengths.  The lower limit of the pair
spectra is at $\epsilon_a/(1 + a^2)^{1/2}$, which is independent of
$B$, provided that $a \gg 1$.  The upper limit is a fixed multiple of
this value, and so remains constant as well.

\section{Semi-numerical model}

The analytic models of Paper I did not take into account the variation
of the power-law exponent, $\nu$, with magnetic field.  This does not
affect the calculation of either the curvature or the resonant ICS
cascades, since those mechanisms operate primarily on the primary
photons, but the decline of $\nu$ with $B$ limits the effectiveness of
non-resonant ICS for magnetic fields above $10^{12}$ Gauss.

Allowing for the change of the power law makes the analytic form
impractical, but we can quantify the approximation by using a
simple numerical model.  If we still assume that all photons emitted
by a given emission mechanism are emitted at the peak energy and use
equation (\ref{eq:mult_s_approx} for the spatial range of
pair-production, we can calculate the location of the PFF.

\begin{figure*}
  \plotone{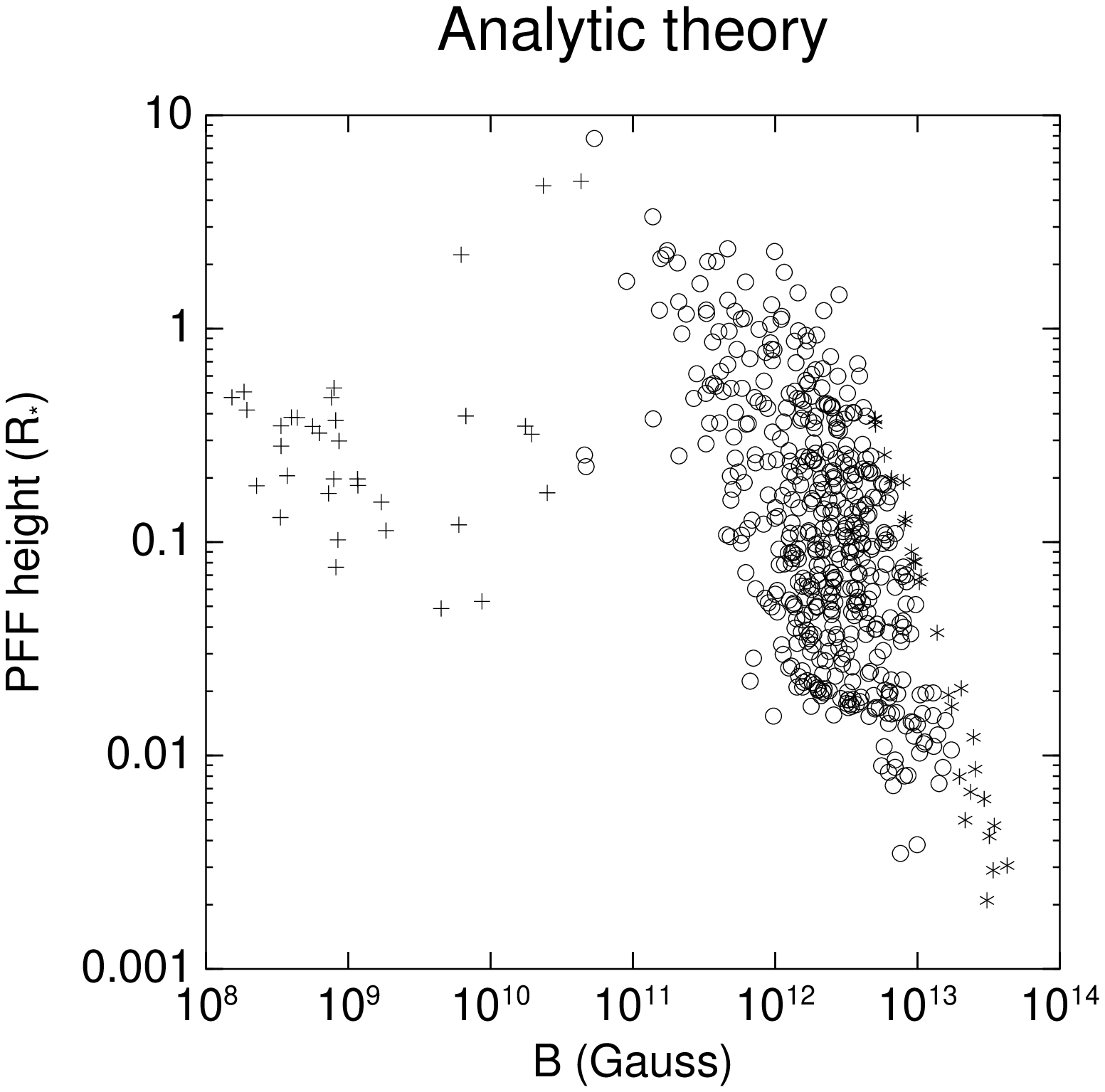}
  \caption{PFF height vs. pulsar magnetic field for the analytic
  theory from Paper I.}
  \label{fig:scatter_anal}
\end{figure*}

\begin{figure*}
  \plotone{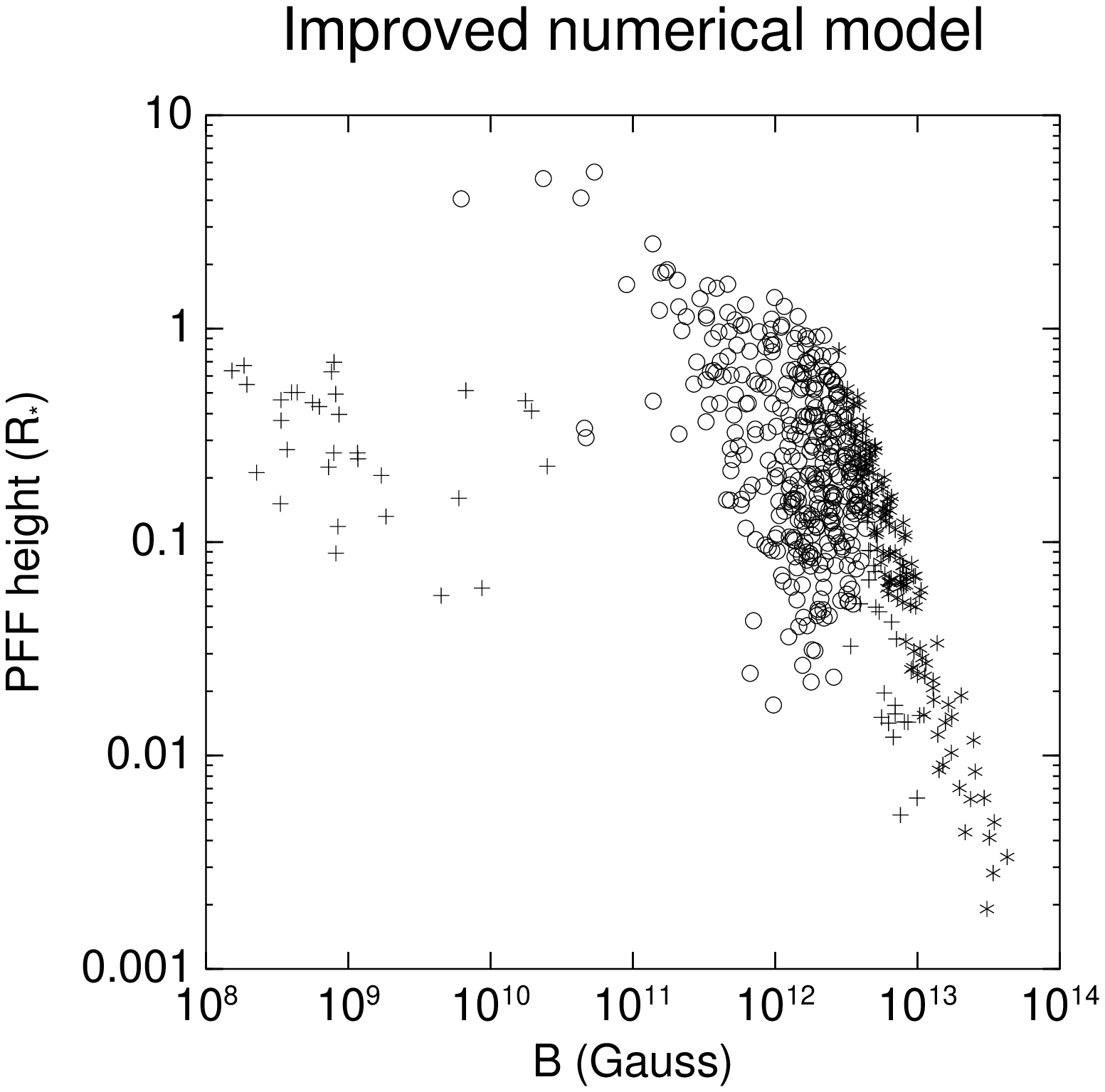}
  \caption{PFF height vs. pulsar magnetic field for the improved numerical
  model.  The numerical
  model matches the analytic at low magnetic field, but reveals an increased
  dominance by RICS at high.}
  \label{fig:scatter_num}
\end{figure*}

In practice, we find that this method yields results which are nearly
indistinguishable from the full numerical method, in the regime where
the full numerical method predicts a finite PFF height.  In Figure
\ref{fig:scatter_anal}, we show the calculated PFF heights from the
analytic model and in Figure \ref{fig:scatter_num} those from the
semi-numerical model.

\section{Full cascade}

The same numerical procedure used to calculate the expected response
to one photon may be applied to a full model of a pulsar polar cap.
We will ignore any variation across the polar cap, concentrating on a
typical field line instead.

Given the pair production physics discussed above, a polar cap model
only requires knowledge of the electric potential accelerating
particles off the surface of the star, the relevant emission
mechanisms which inject high-energy photons into the magnetosphere,
and a model for the temperature of the star.

As a model of the accelerating potential, we use the cubic
approximation to the \citet{muslimov92} GR acceleration derived in
Paper I.  This model is easily computable and sufficiently accurate
for our purposes.

The emission mechanisms included are curvature radiation and inverse
Compton scattering, both resonant and non-resonant.  The spectrum of
curvature radiation is well-known, c.f. \citet{jackson75}, while we
use equation (\ref{eq:n_nrics}) for non-resonant scattering,
with spatial dilution terms included, and equation (\ref{eq:n_rics})
derived from \citet{dermer90} for resonant scattering.  See Appendix
\ref{sec:emission_mechs} for descriptions of the precise emissivities
used.

The stellar temperature is the most uncertain quantity in these detailed 
models. Not only is the precise cooling rate for neutron stars still a 
topic of debate, but the polar cap is likely heated by beam particles 
reversed and accelerated back down onto the cap. To minimize the 
uncertainties, we simply assume a fixed stellar temperature of $10^6$ K.

From the emission mechanisms, we find the Lorentz factor of the beam
as a function of altitude, and may then inject the appropriate
emission spectrum at each radial bin in our grid.  To ensure energy
conservation, we slightly adjust the peak amplitude of each injected
spectrum to compensate for the effects of the discrete energy grid.

We run the model multiple times.  First, given an assumed polar cap
temperature, we run it as described above to find the pair formation
front altitude, namely the altitude at which $\kappa_g$ pairs have
been produced per primary.  At this point the pair plasma is
sufficiently dense to short out the accelerating electric field.

We then recompute the Lorentz factor of the beam, with acceleration
halting at the PFF, inject the new spectra, and follow the cascade to
produce the final output $\gamma$-ray and pair spectra.

\begin{figure*}
  \plotone{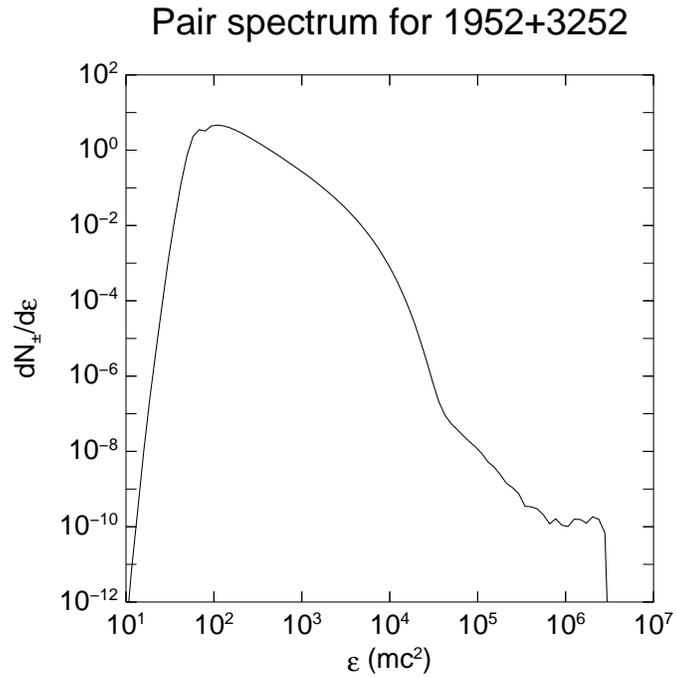}
  \caption{Expected pair spectrum for J1952+3252.}
  \label{fig:pairs_1952}
\end{figure*}

\begin{figure*}
  \plotone{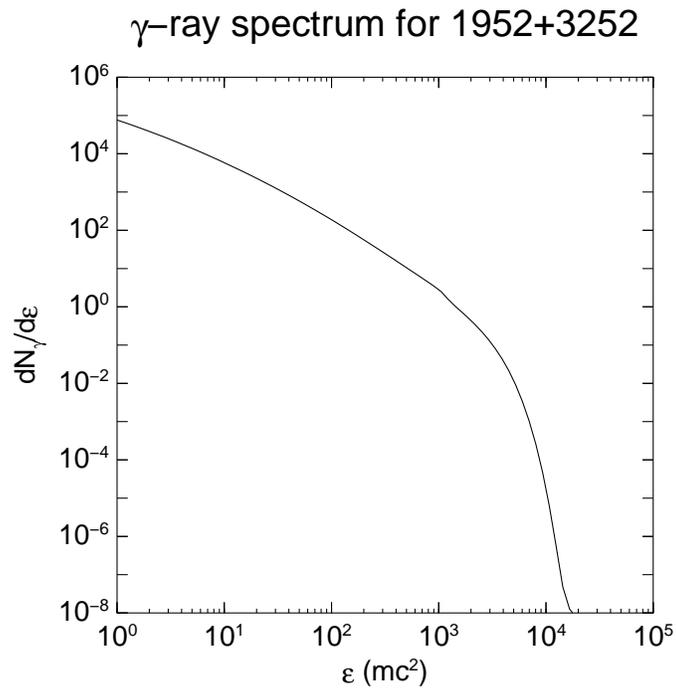}
  \caption{Expected gamma-ray spectrum for J1952+3252, showing the
    expected $\nu=-3/2$ dependence.}
  \label{fig:gamma_1952}
\end{figure*}

A typical output pair spectrum is shown in Figure
\ref{fig:pairs_1952}, and a typical output $\gamma$-ray spectrum in
Figure \ref{fig:gamma_1952}.  These show many of the characteristics
of the single-photon response.  The pair spectrum shows the rapid
rise, $\nu = -3/2$ power-law, and exponential tail, but it also shows
a low-amplitude tail of high-energy particles, due to the effects of
inverse Compton scattering.

For most pulsars, the final pair and $\gamma$-ray spectra are similar
to those in Figures (\ref{fig:pairs_1952}) and (\ref{fig:gamma_1952}).

\begin{figure*}
  \plotone{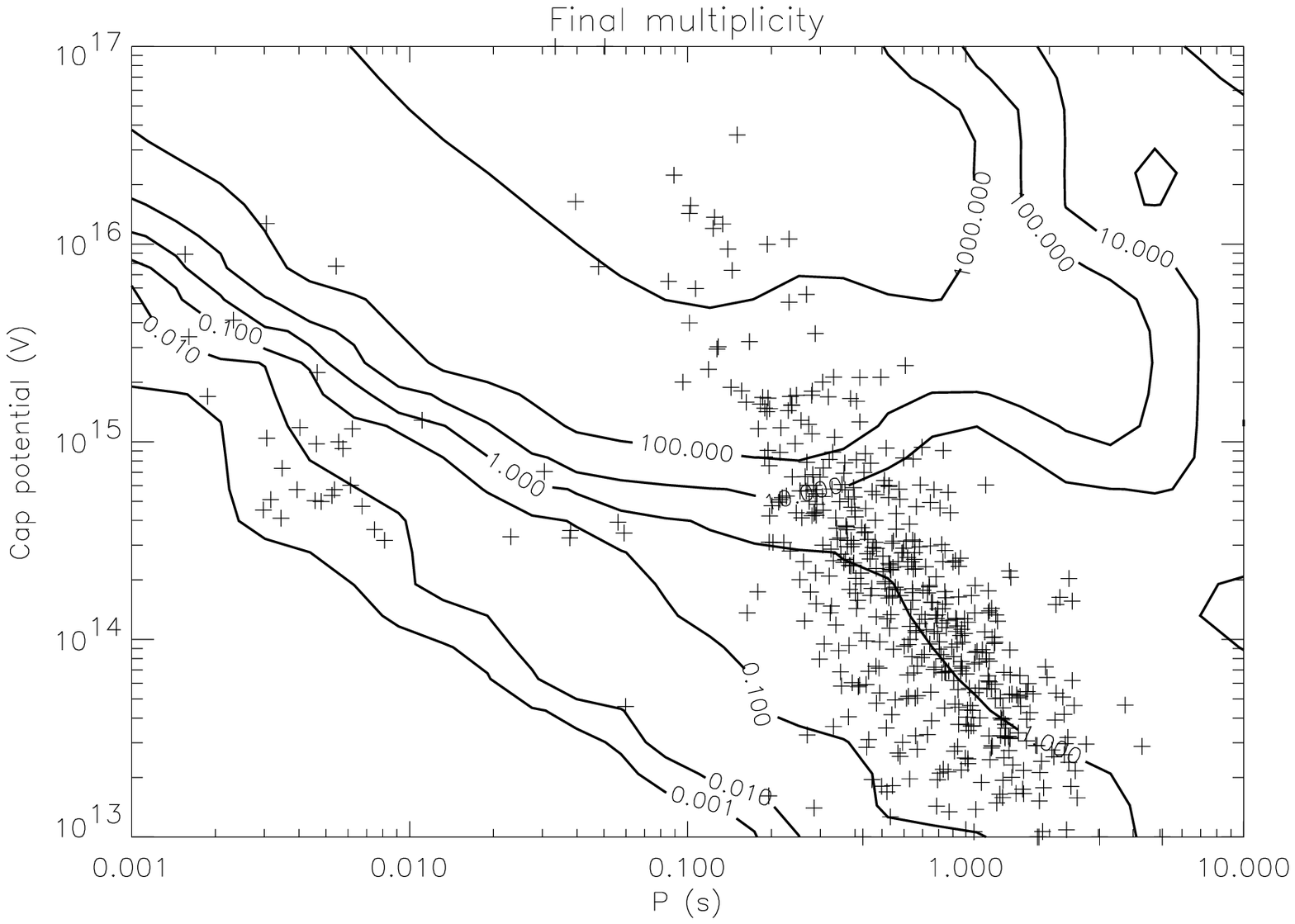}
  \caption{Contour plot of expected multiplicity vs. P, $\Phi$, for a
    fixed stellar temperature of $10^6$ K.}
  \label{fig:contour}
\end{figure*}

\begin{figure*}
  \plotone{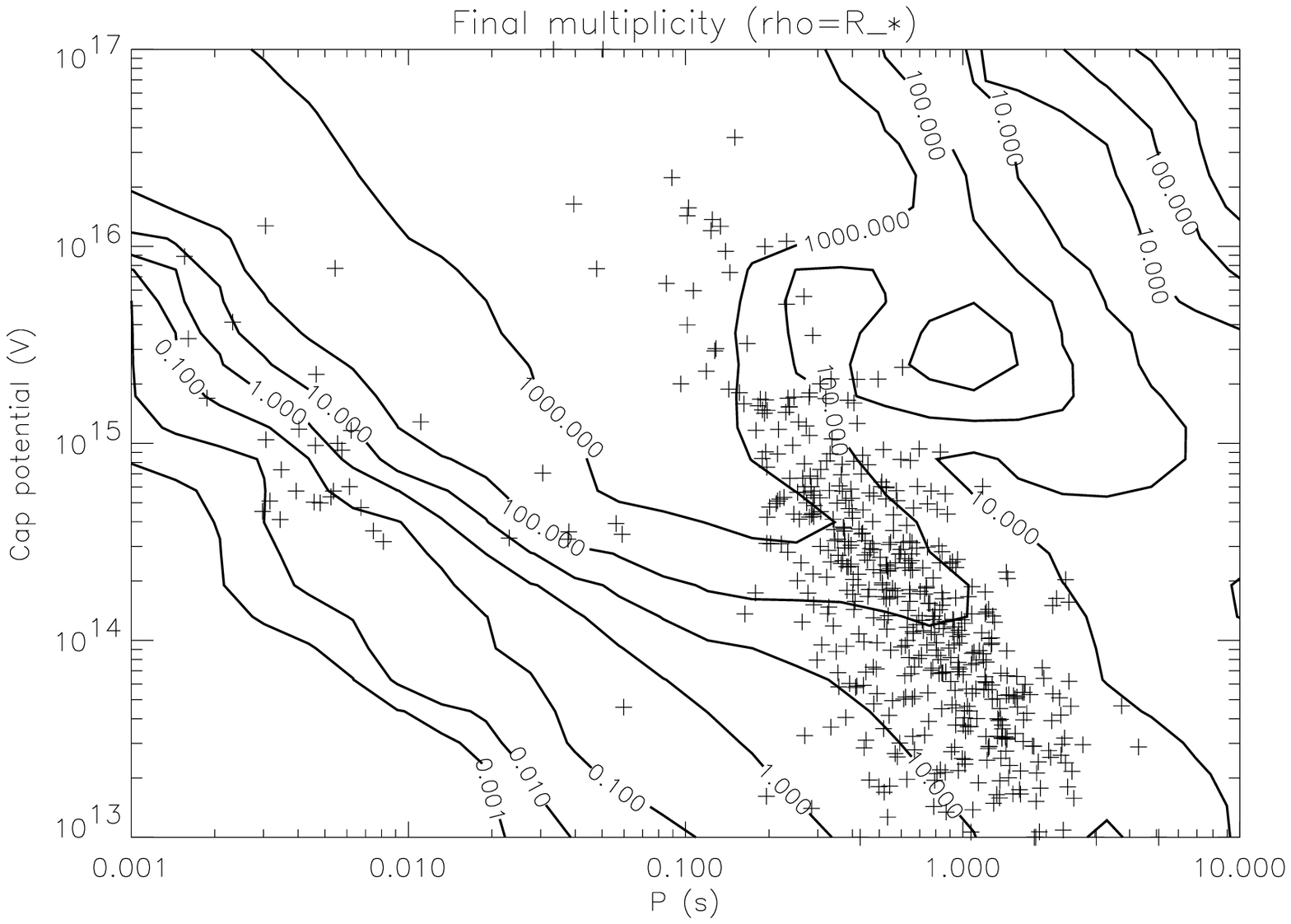}
  \caption{Contour plot of expected multiplicity vs. P, $\Phi$, for a
    fixed stellar temperature of $10^6$ K, with a field line radius of
    curvature, $\rho$, equal to the stellar radius.}
  \label{fig:contour_rho}
\end{figure*}

In Figure \ref{fig:contour}, we plot the multiplicity as a function
of pulsar period and cap potential for a fixed stellar temperature of
$10^6$ K.  Given that fixed temperature, most pulsars have multiplicities
ranging from 0.1--100, while most of the millisecond pulsars have lower
multiplicities.  In Figure \ref{fig:contour_rho}, we set the radius of
curvature of the field lines equal to the stellar radius, in order to
model roughly the effects of an offset dipole.  Here, most pulsars have
multiplicities in the range of 10--1000, while the MSPs remain with
significantly lower multiplicities.  However, a parallel offset of the dipole
moment would also increase the surface field, effectively increasing the
cap potential.  Given the sharp gradient in multiplicities at low-$P$,
this could easily bring the multiplicities of the MSPs into parity with
those of the other pulsars.

\section{Conclusions}

This paper has given several approximations and descriptions of the 
pair creation process in pulsars, in the hopes that they will be useful
for other researchers examining these objects.  The development of the
pair cascade in space is deceptively simple, a fact which has been used
in Paper I to obtain several useful results about the regimes of dominance
of the various emission mechanisms.  A more detailed model, maintaining
the variability of
the power-law exponent, produces qualitatively identical results.  Many
pulsars have their PFF set by non-resonant ICS, with curvature dominating at
high potential, and resonant ICS at high magnetic field.

In addition, due to the effects of NRICS, many pulsars seem to operate
with comparatively low pair multiplicities.  The total $\kappa$ is in
the range of 1--10 more often than it is in the 1000's predicted by
curvature models.

The OTS approximation to the pair cascade describes the spectrum of
$\gamma$-rays and particles produced by the pair production process very
well, although the OTS pair spectrum underestimates the number of low
energy pairs, due to the decline of the magnetic field with altitude.

The full cascade model may be used to predict the $\gamma$-ray and pair
spectra of individual pulsars.  Although the pair spectrum produced
will be modified by RICS, the high-energy $\gamma$-ray spectrum predicted
should be observable.  The $\gamma$-ray output from pulsars will be
the subject of a subsequent paper.

\acknowledgments

We wish to acknowledge the assistance of Alice Harding, who provided
the data for the comparisons of Figures \ref{fig:harding_gamma} and
\ref{fig:harding_pairs}.

\appendix

\section{Emission mechanisms}
\label{sec:emission_mechs}

For convenience, we describe the forms of the various emission mechanisms
used in the numerical calculations.

\subsection{Curvature emission}

For curvature emission, we use the standard formula,
\begin{equation}
  \frac{\partial^2 N_{c}}{\partial t \partial \epsilon_1} = 
    \frac{\sqrt{3}}{3 \pi} \frac{\alphaF c}{\lambdaC}
    \frac{1}{\gamma^2} \int_{\epsilon/\epsilon_c}^\infty
    dx \, K_{5/3}(x)
\end{equation}
where $\epsilon_c = (3/2) \lambdaC \rho \gamma^{-3}$, and $K_{5/3}$ is
the modified Bessel function of order $5/3$.

\subsection{Resonant ICS}

For resonant ICS, we use the result from \citet{dermer90},
\begin{equation}
  \frac{\partial^2 N_{r}}{\partial t \partial \epsilon_1} =
    \frac{1}{2} \frac{\alphaF c}{\lambdaC}
      \frac{T}{\gamma^{3} \beta^{2}}
      \ln \frac{\gamma T \Delta \mu}{\eb}
      H \left(\epsilon_{1}, \frac{\eb}{2 \gamma}, 2 \gamma \eb 
      \right)
  \label{eq:n_rics}
\end{equation}
where $H$ is the top hat function, defined before.

\subsection{Non-resonant ICS}

Although the curvature emission and resonant inverse-Compton scattering
spectra are well-represented in the literature, the non-resonant ICS
spectrum is not, to the accuracy we desire.

The chief difference between the physical situations considered by
\citet{blumenthal70} and conditions in the pulsar magnetosphere is
that the background X-rays in pulsars are not isotropic.  They
originate from the hot surface of the star itself and only occupy a
range in $\theta$ from $\theta=0$ to at most $\theta=\pi$.  In terms
of $\mu=\cos \theta$, thermal photons are emitted with angles from 1
to $\mu_c$, where
\begin{equation}
  \mu_c = \frac{z}{\sqrt{a^2 + z^2}}
\end{equation}
where $a$ is the radius of the polar cap, typically $a=\theta_c R_*$,
and $z$ is the height above the stellar surface.  For simplicity, we
only consider emission along the $z$-axis, where the incoming photons
are symmetric in $\phi$.  The range of $\mu$ populated with photons is
then
\begin{equation}
  \Delta \mu = 1 - \mu_c = 1 - \frac{z}{\sqrt{a^2 + z^2}}.
\end{equation}

For the isotropic case, $\Delta \mu = 2$.  If $\Delta \mu < 2$, the
simplest consequence is that there are fewer photons, by a factor of
$\Delta \mu / 2$, potentially scattering off of the particle.
Furthermore, these photons are concentrated near $\mu = 1$, so the
mean energy of the photons in the particle rest frame is reduced.

From \citep{blumenthal70}, the full non-resonant inverse Compton cross
section is
\begin{equation}
  \frac{\partial \sigma(\epsilon', \mu')}{\partial \epsilon'_1
    \partial \mu'_1} =
  \frac{1}{2} r_0^2 \left(\frac{\epsilon'_1}{\epsilon'}\right)^2
     \left(\frac{\epsilon'}{\epsilon'_1} + \frac{\epsilon'_1}{\epsilon'}
      - (1 - \mu^{\prime 2}_1) \right)
  \delta (\epsilon'_1 - \frac{\epsilon'}{1 + (\epsilon' / mc^2)
    (1 + \mu'_1)} )
\end{equation}
The primed frame is the particle rest frame, the unprimed frame is fixed
to the neutron star itself, and the subscript $1$ indicates the scattered
particles and angles.  We have made the approximation that all of the
incoming photons in the particle rest frame are beamed down the $z$-axis,
so that the angle between the incoming and the scattered photon is simply
equal to the latitude of the scattered photon, $\Theta' = \pi - \theta'_1$.
This approximation works very well for large Lorentz factors, but is
less accurate in the Thomson regime.  However, it gives negligible
error in the Klein-Nishina regime.

The rest-frame scattered number emissivity is then
\begin{equation}
  j'(\epsilon'_1, \mu'_1) = 2\pi n' \int d\epsilon' d\mu' \,
   \frac{\partial \sigma(\epsilon', \mu')}{\partial \epsilon'_1
     \partial \mu'_1}
   I'(\epsilon', \mu').
\end{equation}
Using the $\delta$-function in the cross section to perform the
$\epsilon'$ integral yields
\begin{equation}
  j'(\epsilon'_1, \mu'_1) = \frac{1}{2} n' r_0^2
    \left(\frac{\epsilon_0'}{\epsilon'_1} + \frac{\epsilon'_1}{\epsilon_0'}
     - (1 - \mu^{\prime 2}_1) \right)
  2 \pi \int d\mu' \, I'(\epsilon'_0, \mu')
\end{equation}
where
\begin{equation}
  \epsilon'_0 \equiv \frac{\epsilon'_1}{1 - (\epsilon'_1 / mc^2)(1
    + \mu'_1)}
\end{equation}

The integral over the source intensity is
\begin{equation}
     2 \pi \int_{-1}^{1} d\mu' \, I'(\epsilon'_0, \mu') = 2 \pi
     \int_{-1}^{1} d\mu' \, I(\epsilon(\epsilon'_0, \mu'), \mu(\epsilon'_0, 
     \mu')) \left(\frac{\epsilon'_0}{\epsilon(\epsilon'_0, \mu')}\right)^2.
\end{equation}
Changing variables to $\epsilon$, using $\epsilon = \gamma \epsilon'_0
(1 + \beta \mu')$, this becomes
\begin{equation}
  2 \pi \int_{-1}^{1} d\mu' \, I'(\epsilon'_0, \mu') =
    2 \pi \frac{\epsilon'_0}{\gamma \beta}
    \int_{\epsilon_{min}}^{\epsilon_{max}} d\epsilon \, \epsilon^{-2} 
    I(\epsilon, \mu(\epsilon))
\end{equation}

The limits of the $\epsilon$ integration are set by which energies may
be scattered such that they have energy $\epsilon'_0$ in the rest
frame.  The minimum possible energy is $\epsilon_{min} =
\epsilon'_0/\gamma (1 - \beta \mu_c) \approx \epsilon'_0/\gamma \Delta
\mu$, while the maximum is $\epsilon_{max} = \epsilon'_0/\gamma(1 -
\beta) \approx 2 \gamma \epsilon'_0$, where the approximations assume
that $\beta \approx 1$.

The source function we adopt is that of a black-body, emitting at
temperature T in units of $mc^2$, into a range of solid angle $\delta
\mu = 1 - \mu_c$.  This corresponds to a specific intensity of
\begin{equation}
    I(\epsilon, \mu) = \frac{1}{4 \pi^3} \frac{c}{\lambdaC^3} 
      \frac{\epsilon^2}{\exp(\epsilon/T) - 1} H(\mu, \mu_c, 1)
\end{equation}
where $H(x, a, b) = 1$ for $a < x < b$, and $0$ otherwise.

Combining all these components, we find
\begin{eqnarray}
    2 \pi \int_{-1}^{1} d\mu' \, I'(\epsilon'_0, \mu') & = &
      \frac{1}{2 \pi^2} 
      \frac{c}{\lambdaC^3} \frac{\epsilon'_0}{\gamma \beta} 
      \int_{\epsilon_{min}}^{\epsilon_{max}} 
      \frac{1}{\exp(\epsilon/T) - 1} \\
    & = & \frac{1}{2 \pi^2} \frac{c}{\lambdaC^3} \frac{\epsilon'_0 T}{\gamma 
      \beta} \ln \left(\frac{1 - \exp(-2 \gamma \epsilon'_0/T)}{1 - 
      \exp(-\epsilon'_0/\gamma T \Delta \mu)}\right)
      \label{eq:I_integral}
\end{eqnarray}

The numerator in the log is equivalent to 1 for all energies with
significant emission.  We then have, 
\begin{eqnarray*}
  N(\epsilon_1) & = & 2\pi \frac{1}{n'} \int d\epsilon'_1 d\mu'_1 \,
     \frac{1}{\gamma^2 (1 + \beta \mu'_1)}
     \delta(\epsilon'_1 - \frac{\epsilon_1}{\gamma (1 + \beta \mu'_1)})
     j'(\epsilon'_1, \mu'_1) \\
  N(\epsilon_1) & = & \frac{1}{2\pi} \frac{\alphaF^2 c}{\lambdaC}
    \frac{T}{\gamma^2 \beta} \int  d\epsilon'_1 d\mu'_1 \,
    \frac{-1}{1 + \beta \mu'}
    \delta(\epsilon'_1 -
      \frac{\epsilon_1}{\gamma (1 + \beta \mu')}) \times \\
  & & \quad
    \left(\frac{\epsilon'_0}{\epsilon'_1} +
    \frac{\epsilon'_1}{\epsilon'_0}
      - (1 - \mu^{\prime 2}_1) \right) \epsilon'_0
      \ln (1 - \exp(-\epsilon'_0/ \gamma T \Delta \mu))
\end{eqnarray*}
where we have substituted $r_0^2 / \lambdaC^{3} = \alphaF^2/\lambdaC$.

If we then assume that $\beta \approx 1$ so $\epsilon'_1/\epsilon'_0 =
1 - \epsilon_1/\gamma$, and evaluate the $\epsilon'_1$ integral,
we obtain
\begin{eqnarray}
  \frac{\partial^2 N_{c}}{\partial t \partial \epsilon_1} & = & \frac{1}{2 \pi} \frac{\alphaF^2 c}{\lambdaC}
    \frac{\epsilon_1 T}{\gamma^4 \beta (1 - \bar{\epsilon_1})} 
    \int d\mu'_1 \, \frac{-1}{(1 + \beta \mu'_1)^2}
    \left( (1 - \bar{\epsilon_1}) + \frac{1}{1 -
    \bar{\epsilon_1}} - 1 + \mu^{\prime 2}_1\right) \times \nonumber \\
  & & \quad \ln \left(1 - \exp
    \left( \frac{-\epsilon_1}{\gamma^2(1 + \beta \mu'_1) (1 -
    \bar{\epsilon_1}) T \Delta \mu} \right) \right)
\end{eqnarray}
where we have defined $\bar{\epsilon_1} \equiv \epsilon_1/\gamma$.

Changing variables to
\begin{equation}
  y = \frac{\epsilon_1}{\gamma^2(1 + \beta \mu'_1) (1 -
    \bar{\epsilon_1}) T \Delta \mu},
\end{equation}
gives, after setting $\beta = 1$,
\begin{equation}
  N(\epsilon_1) = \frac{1}{2 \pi} \frac{\alphaF^2 c}{\lambdaC}
    \frac{T^2 \Delta \mu}{\gamma^2 (1 - \bar{\epsilon}_1)}
    \int dy \, 
    \left(1 +  (\mu^{\prime 2}_1 - 1) (1 - \bar{\epsilon}_1) + (1 -
      \bar{\epsilon}_1)^2 \right)  \ln (1 - e^{-y}).
\end{equation}

Approximating this further by replacing $\mu_1^{\prime 2}$ with its
average over the range, $\average{\mu_1^{\prime 2}}$, and performing the
integral yields
\begin{equation}
  N(\epsilon_1) = \left. \frac{1}{2\pi} \frac{\alphaF^2 c}{\lambdaC}
    \frac{T^2 \Delta \mu}{\gamma^2 (1 - \bar{\epsilon_1})}
    \left( 1 + (\average{\mu_1^{\prime 2}}-1) (1 - \bar{\epsilon_1}) + (1 -
      \bar{\epsilon_1})^2 \right) \mbox{Li}_2(e^{-y})
    \right|_{y_{max}}^{y_{min}}
    \label{eq:n_nrics}
\end{equation}
where
\begin{eqnarray}
  \int dy \ln(1 - e^{-y}) & = & \mbox{Li}_2(e^{-y}) \\
  y_{min} & = & \frac{\epsilon_1}{2 \gamma^2 (1 - \bar{\epsilon_1}) T
  \Delta \mu} \\
  y_{max} & = & \frac{\epsilon_1}{(1 - \bar{\epsilon_1}) T \Delta \mu}
\end{eqnarray}
where $\mbox{Li}_2(x)$ is the dilogarithm function, defined as
\begin{equation}
  \mbox{Li}_n(x) = \sum_{i=1}^{\infty} \frac{x^i}{i^n}.
\end{equation}
The dilogarithm, $\mbox{Li}_2(x) = x$ for small $x$, and
$\mbox{Li}_2(1) = \pi^2/6$.

The angle average, $\average{\mu_1^{\prime 2}}$, clearly lies between 0 and 1.
Its value is given by
\begin{equation}
  \average{\mu_1^{\prime 2}} = \frac{\int_{z_0}^{z_1} dz \, 
    (z-1)^2 z^{-2} (- \ln (1 - \exp(-\kappa / z)))}
  {\int_{z_0}^{z_1} dz \, z^{-2}
     (- \ln (1 - \exp(-\kappa / z)))}
\end{equation}
where
\begin{eqnarray}
  \kappa & \equiv & \frac{\bar{\epsilon_1}}{\gamma \Delta \mu T
    (1 - \bar{\epsilon_1})}     \\
  z & \equiv & 1 + \beta \mu'_1 \\
  z_0 & = & 1 - \beta \approx 0 \\
  z_1 & = & 1 + \beta \approx 2
\end{eqnarray}
Adopting the approximate forms of $z_0$ and $z_1$ does not perceptibly
change the value of the integral, but reduces $\average{\mu_1^{\prime
2}}$ to being a function only of the single parameter $\kappa$. A good
approximation to the value of $<\mu^{\prime 2}>$ is
\begin{equation}
  \average{\mu_1^{\prime 2}} \approx 1 - 0.76 \exp(-(\ln \kappa)^2/10)
\end{equation}
For the computation, we simply tabulate $\average{\mu_1^{\prime 2}}$.

The power emitted by the scattering has two asymptotic limits.  If
$\gamma T \Delta \mu <  1$, then emitted power matches the Thomson
value,
\begin{equation}
  P^{(Th)}_n = \frac{4 \pi^3}{135} \frac{\alphaF^2 c}{\lambdaC} \gamma^2
    \Delta \mu^3 T^4,
\end{equation}
while if $\gamma T \Delta \mu > 1$, the emitted power is approximately
\begin{equation}
  P^{(KN)}_n \approx \frac{\pi}{12} \frac{\alphaF^2 c}{\lambdaC} T^2 \Delta \mu
    \ln 2 \gamma T \Delta \mu
\end{equation}
which matches the Klein-Nishina limit of \citet{blumenthal70} when
$\Delta \mu = 2$.  The transition between the two limits is purely a
function of $\gamma T \Delta \mu$.  The emitted power is
\begin{equation}
  P_{n} = \frac{1}{2 \pi} \frac{\alphaF^2 c}{\lambdaC} T^2 \Delta \mu
    f(\gamma T \Delta \mu)
\end{equation}
Where $f$ is a numerically calculated function which has asymptotic limits
\begin{equation}
  \begin{array}{rcll}
    f(x) & \approx & \frac{8 \pi^4}{135} x^2, & x \ll 1 \\
    f(x) & \approx & \frac{\pi^2}{6} \log 2 x, & x \gg 1.
  \end{array}
\end{equation}

\end{document}